\documentclass[aps,prd,twocolumn,showpacs,superscriptaddress]{revtex4-2}
\usepackage{amsmath}
\usepackage{amsfonts}
\usepackage{graphicx}
\usepackage{hyperref}
\usepackage{color}
\usepackage{tikz}
\usepackage{caption} 
\begin{document}

\title{Accretion with Back-Reaction onto Cylindrically Symmetric Black Hole with Energy Conditions Analysis}
\author{M. Zubair Ali Moughal}
\email{moughalzubair@gmail.com}
\affiliation{Department of Basic Sciences and Humanities,\\
 College of Electrical and Mechanical Engineering,\\
 National University of Science and Technology,\\
		Islamabad, Pakistan}
\author{Kamran Qadir Abbasi}
\email{kamranqadir@numl.edu.pk}
\affiliation{Department of Mathematics\\
	Faculty of Engineering and Computing (FE\&C)\\
	National University of Modern Languages (NUML),\\
	Sector H-9, Islamabad, Pakistan.\\}
\date{\today}

\begin{abstract}
This paper is devoted to study back-reaction effects from matter accretion onto a cylindrically symmetric black hole using a perturbative scheme, focusing on cases where accretion reaches a quasi-steady state. We examine three distinct models by deriving corrections to the metric coefficients and obtaining expressions for the mass function. We analyze energy conditions, the self-consistency of the corrected solution and present formulas for the corrected apparent horizon and discussed thermodynamic properties. Our results align with the Vaidya form near the apparent horizon, regardless of the energy-momentum tensor's form. Furthermore, we show that for a charged cylindrically symmetric black hole, the corrected mass term resembles that of the static case, indicating that charge does not alter the corrected metric form in this perturbative approach.
\end{abstract}


\maketitle

\section{Introduction}
Cylindrical black holes (BHs) or black strings (BS) are the cylindrically symmetric static solutions of the Einstein-Maxwell field equations (EMFEs) together with a negative cosmological constant. The study of BS has along history. The first exact solution for BS was discovered by J. P. Lemos in \cite{rl}. Following this initial work, R. G. Cai and Y. Zhang calculated the charged version \cite{2}. Shortly thereafter, solutions for the rotating and rotating charged versions of the BS were also found \cite{3}.

Around a BS, there is an accretion disk. This accreting matter falls into the BS, termed as a back-reaction of accreting matter onto a BS. In Newtonian gravity, first the problem of matter accretion onto compact objects has been formulated in a self similar manner by Bondi \cite{4}. The concept of matter accretion onto BHs traces back to the 1970s \cite{r5}, where pioneering work by Shakura and Sunyaev (1973)  introduced the ``disk model" for accretion flows around compact objects, including BHs \cite{r6}. This disk model emphasized how gas spirals into BHs, releasing energy and potentially affecting the BH's growth rate and observable properties. Shakura and Sunyaev's \cite{r6} approach set a foundational framework for understanding how accretion disks contribute to the observable luminosity of BHs and led to the recognition of accretion as a fundamental mechanism in BH evolution and active galactic nuclei (AGN) theory \cite{r7,r8}.
 
Following these early models, the idea of back-reaction, or the impact of accreting matter on the BHs own properties, was explored \cite{r9}. In the early 1980s, seminal work by Bardeen and Wagoner\cite{r10} examined the effects of angular momentum and energy transfer during accretion, proposing that the BHs spin and mass could be incrementally modified through this interaction \cite{r11}. This line of inquiry laid the groundwork for understanding back-reaction as not merely a passive process but as one where the accreting matter might influence the spacetime geometry near the BH, setting up a feedback mechanism that depends on the accretion flows nature \cite{r12}.

In recent years, with advancements in perturbative methods and numerical relativity, the study of BH accretion has evolved to include back-reaction effects \cite{r13} in more complex spacetimes. Modern treatments often apply perturbative methods \cite{r14} to acquire how small amounts of infalling matter influence the metric around BHs, allowing for a more detailed analysis of back-reaction effects on rotating and non-rotating BHs alike. Contemporary studies by Babichev et al. (2018) \cite{r15} and others highlight how accretion with a focus on energy conditions has extended to cosmological contexts, opening up new applications and insights in high-energy astrophysics \cite{r16}.

In this paper, we investigate the back-reaction effects using a perturbative scheme onto the BS solution. Initially, we analyze the static BS case, approximating the mass at zeroth order. This results in a mass expression as a function of time \(t\) and radial coordinate \(r\), which we refer to as the {\it running mass}. We then examine the self-consistency of the solution and  energy conditions analysis using the mass function. Furthermore, we explore three different accretion models, providing graphical insights based on the running mass. By considering the perturbed metric, we derive the expression for the corrected apparent horizon and investigate how accreting matter affects the horizon over time, as well as the energy density and pressure in these models. We also discuss the thermodynamics of the corrected  horizon by calculating temperature and entropy expressions for each case. Additionally, we show that for a charged BS, the mass function expression resembles that of the static case, without the charge contributing to the  mass in the perturbative approach. Hence, this approach demonstrates that the accretion back-reaction onto both charged and uncharged cylindrically symmetric BS solutions is identical.

Babichev et al. \cite{r22} applied a similar approach to study spherically symmetric static BHs. In our work, we extend this method to the case of BS. Specifically, we employ this framework to analyze both charged and uncharged BS and compare the results to explore how back-reaction affects their properties. Additionally, we provide a detailed analysis of the thermodynamics of the corrected BS metric for both cases. 

Notably, we compare the corrected mass with the ADM mass and derive the relationship between them, highlighting their interdependence. We believe this constitutes a valuable contribution to the available literature on the study of BS.

The structure of this paper is as follows:
In the next section, we present the mathematical formalism necessary to investigate the back-reaction effects. In Section III, we apply the mathematical formalism to a static BS and derive the corrected  mass term. Section IV addresses the self-consistency of  solution and energy conditions analysis, providing a physical interpretation of the back-reaction phenomena. Section V discusses the analysis for different accreating models. In Section VI, we examine the back-reaction effects for a charged BS, and the paper concludes with discussion and conclusion in Section VII.

\section{Perturbation Scheme}
	In the study of BH accretion, a prevalent assumption is the neglect of back-reaction effects due to the typically negligible mass of the accreting matter compared to the BH's mass \cite{r17}. This approximation allows for simpler analytic models, but it overlooks the complex interplay between the gravitational field of the BH and the dynamics of the infalling matter. The challenge of solving the equations governing accreting matter with back-reaction arises from the intricate coupling between matter and spacetime geometry \cite{r18}, leading to non-linear effects that are difficult to analyze. Consequently, few analytic solutions have been derived that fully account for these back-reaction phenomena \cite{r19}. In this section, we will investigate the implications of back-reaction effects due to matter accretion, aiming to enhance our understanding of their role in the dynamics of BS.
	
	The complete solution of the EFEs in the general case of
	BH accretion is still unknown but there are some special cases of the
	energy-momentum tensor for which the exact solution were found i.e. the
	Vaidya and Tolman solutions \cite{r20, r21}. In this Section we review the
	peterbative approach to find the solution. This approach was also used by
	Babichev et al. \cite{r22}. Here we correct the metric due to accreting matter
	having energy-momentum tensor at zeroth-order approximation (i.e.
	back-reaction is neglected) and we find the first order correction of metric with back-reaction. Mathematically, we begin with the EFEs with metric tensor $g_{\mu v}$ and $\varphi$ represents the degree of freedom associated with accreting matter,
\begin{equation}
G_{\mu v}\left[g_{\mu v}\right]=8 \pi T_{\mu v}\left[g_{\mu v}, \varphi\right],
\label{0}
\end{equation} 
and the equations of motion for matter field are $E\left[g_{\mu v}, \varphi\right]=0$, which can also derived form the Bianchi identities. In Eq. (\ref{0}), if we neglect the back-reaction, we obtain a zeroth-order approximation. Specifically, in this approximation, the solution for the metric is the vacuum solution,i.e $g_{\mu v}^{(0)}=g_{\mu v}^{v a c} $ so that $G_{\mu v}\left[g_{\mu v}^{(0)}\right]=0 $. Also in the same zeroth order approximation, the solution for the equation for matter field(s), $\varphi^{(0)}$, is computed as
\begin{equation}
E\left[g_{\mu v}^{(0)}, \varphi^{(0)}\right]=0.
\label{00}
\end{equation}
Now, to proceed with the first-order approximation, we substitute \( g_{\mu\nu}^{(0)} \) and \( g_{\mu\nu}^{(1)} \) as follows: On the right-hand side of Eq. (\ref{0}), we substitute \( g_{\mu\nu}^{(0)} \), and on the left-hand side, we substitute \( g_{\mu\nu}^{(0)} + g_{\mu\nu}^{(1)} \). This leads to the EEFs taking the form,
\begin{equation}
	G_{\mu\nu}\left[ g_{\mu\nu}^{(0)} + g_{\mu\nu}^{(1)} \right] = 8\pi T_{\mu\nu}\left[ g_{\mu\nu}^{(0)} \right]. \label{1}
\end{equation}
	Also, we assumed that \( g_{\mu\nu}^{(0)} \) is greater than \( g_{\mu\nu}^{(1)} \). Next, we apply this on the static BS metric.
\section{Static Black String}
The static, uncharged BS metric with a negative cosmological constant \(\alpha^2 = -\frac{\Lambda}{3} > 0\) in anti-de Sitter \cite{r23} spacetime is given by 
\begin{align}
ds^{2}=&-\left(  \alpha^{2}r^{2}-\frac{m_{0}}{r}\right)  dt^{2}+\dfrac
{1}{\left(  \alpha^{2}r^{2}-\dfrac{m_{0}}{r}\right)  }dr^{2} \nonumber\\
&+r^{2}d\theta
^{2}+\alpha^{2}r^{2}dz^{2}. \label{2}%
\end{align}
This metric represents the solution at the zeroth-order approximation. To include back-reaction effects, we introduce a perturbed EFEs as defined in Eq. (\ref{1}), the function \(m(t, r)\) replaces the constant \(m_0\). At zeroth order, \(m(t, r)\) reduces to \(m_0\), where \(m_0\) represents the mass without the back-reaction effect and thus remains constant. Simplifying Eq. (\ref{2}),
	\begin{align}
	ds^{2}=&-\left(  \alpha^{2}r^{2}-\frac{m(t,r)}{r}\right)  dt^{2}+\dfrac
	{1}{\left(  \alpha^{2}r^{2}-\dfrac{m(t,r)}{r}\right)  }dr^{2}\nonumber\\
	&+r^{2}d\theta
	^{2}+\alpha^{2}r^{2}dz^{2}. \label{3}%
	\end{align}
  The components of the Einstein tensor are expressed as follows:
	\begin{align}
	\label{04}G_{0}^{0}&=G_{1}^{1}=\frac{1}{r^{2}}\bigg( 3\alpha^{2}r^{2}- m^{\prime
	} \bigg),\\
	\label{05} G_{0}^{1}&=\frac{\dot{m}}{r^{2}},
	~G_{1}^{0} =\frac{\dot{m}}{\left(  \alpha^{2}r^{3}-m\right)  ^{2}},\\
	\label{06} G_{2}^{2}&=G_{3}^{3}= \frac{1}{2r\left(  -\alpha^{2}r^{3}+m\right) ^{3}}\bigg[
	6m^{3}\alpha^{2}r+\ddot{m}  r^{5}\alpha^{2}\nonumber\\
	&-18\alpha^{4}r^{4}m^{2}-m^{3} m^{\prime \prime}  
 -\ddot{m}  r^{2}m+2  \dot{m}  ^{2}r^{2}\nonumber\\
   &-3\alpha^{4}r^{6}m m^{\prime \prime}  +3\alpha^{2}r^{3}	m^{2}  m^{\prime \prime}\nonumber\\
  &+18\alpha^{6}r^{7}m+\alpha^{6}r^{9}  m^{\prime \prime}t
	-6\alpha^{8}r^{10}\bigg].
	\end{align}
We denote derivatives with respect to  $t$ by a dot and those with respect to $ r$ by a prime. By substituting the components of the Einstein tensor into EFEs, we obtain
	\begin{align}
	\label{07}8\pi T_{0}^{0}  &= 8\pi T_{1}^{1} = \frac{1}{r^{2}} \big(3\alpha^{2} r^{2} - m' \big), \\
     \label{08}8\pi T_{0}^{1} &= \frac{\dot{m}}{r^{2}}, ~~8\pi T_{1}^{0} = \frac{\dot{m}}{(\alpha^{2} r^{3} -m)^{2}}, \\
	\label{09}8\pi T_{2}^{2} &= 8\pi T_{3}^{3} = \frac{1}{2r(-\alpha^{2} r^{3} + m)^{3}} [ 6m^{3}\alpha^{2} r + m'' r^{5}\alpha^{2}\nonumber\\
	& - 18\alpha^{4} r^{4} m^{2}  - m^{3} \ddot{m}(t, r) - m'' r^{2} m + 2(m')^{2} r^{2} \nonumber\\
	&- 3\alpha^{4} r^{6} m \ddot{m} + 3\alpha^{2} r^{3} m^{2} \ddot{m}(t, r) + 18\alpha^{6} r^{7} m\\
	&+ \alpha^{6} r^{9} \ddot{m}(t, r) - 6\alpha^{8} r^{10}  ]. \nonumber
	\end{align}
	To analyze the problem effectively, we begin by solving Eq. (\ref{07}) to express \( m^{\prime}(t, r) \) in terms of \( T_{0}^{0} \) and \( T_{1}^{1} \). Subsequently, we solve Eq. (\ref{08}), which provides an expression for \( \dot{m}(t, r) \) in terms of \( T_{1}^{0} \) and \( T_{1}^{1} \). However, it is important to note that not all components of this system of equations are independent. By using the Bianchi identities, we demonstrate that Eq. (\ref{09}) is essentially a linear combination of Eqs. (\ref{07}) and (\ref{08}). Therefore, Eq. (\ref{09}) does not contribute additional information and is not taken into account further. As a result, the expression for the mass is obtained through the solutions of Eqs. (\ref{07}) and (\ref{08}),
	 \begin{equation}
	 m(t,r)=-\int_{r_{0}}^{r}8\pi r^{2}T_{0}^{0}dr+\alpha^{2}r^{3}+m_{0}+8\pi
	 tr^{2}T_{0}^{1}. \label{012}
	 \end{equation}
	Eq. (\ref{012}) represents the central finding of this study. In this context, \( m(t,r) \) denotes the corrected mass, which accounts for the back-reaction of the accreting matter, while \( m_{0} \) refers to the zeroth-order mass, representing the case without back-reaction. The components of the energy-momentum tensor are slowly varying functions of the radial coordinate. From Eq. (\ref{012}), we can express the mass \( m(t, r) \) in the vicinity of the BS horizon as \( r \) approaches \( r_0 \), indicating that near the horizon, the mass can be written as,
	\begin{equation}
	 m(t,r)=-8\pi r^{2}(r-r_{0})|_{r=r_{0}}T_{0}^{0}+\alpha^{2}r^{3}+m_{0}+8\pi
	 tr^{2}T_{0}^{1}. \label{013}%
	 \end{equation}
	 Additionally, for our scheme, Eq. (\ref{013}) is only applicable when the correction is small. Therefore, we require,
	 \begin{align}
	 |8\pi tr^{2}|  &  \ll m_{0}, \nonumber\\
	 \newline|\alpha^{2}r^{3}|  &  \ll m_{0}, \\
	 \newline|-8\pi r^{2}(r-r_{0})|_{r=r_{0}}T_{0}^{0}|  &  \ll m_{0}%
	 .\newline \nonumber
	 \end{align}
	 	\section{Self-consistency of solution and Energy conditions}
	 	
In the context of GR, self-consistency in solutions to the EFEs is fundamental for ensuring the physical validity of any model, especially those involving back-reaction effects and accretion onto BHs \cite{r24}. A self-consistent solution requires that the energy-momentum tensor \( T^{\mu}_ {\nu} \), satisfies the EFEs in a way that aligns with the background metric while remaining dynamically consistent with the boundary conditions or any approximations used \cite{r25}. This approach is commonly found in studies of accreting systems and BHs, where the metric is perturbed by incoming matter \cite{r26}, and any deviations are captured by the resulting back-reaction on spacetime geometry.
	 
In accretion studies, the test-fluid approximation is frequently used to model the dynamics of the accreting matter within a fixed background, generally under the assumption that the accreted mass has minimal impact \cite{r27} on the BH’s mass and metric. This approach is valid under conditions of small energy densities $\rho$ and slow accretion rates $\dot{m}$. The back-reaction can be handled perturbatively. For instance, in works by Michel (1972) \cite{r28} and Babichev et al. (2004) \cite{r29}, the self-consistency of accretion models is maintained by ensuring that the accreting matter's influence remains a small perturbation in comparison to the BHs mass \cite{r30}. 
	 
In our solution, this self-consistency is ensured by establishing small parameters associated with the energy density and accretion rate. By expressing the Einstein tensor components and solving the resulting field equations Eqs.(\ref{04} - \ref{09}), we derive the corrected mass function \( m(t, r) \), which incorporates these small parameters in a manner that respects the background symmetry. Our solution ensures consistency by adhering to the condition \( \rho_{\infty} m^2 \ll 1 \), meaning that the energy density of the accreting matter at infinity is small enough to allow for the test-fluid treatment, where the BS spacetime is only minimally perturbed by the incoming matter.

Furthermore, our framework allows us to relate the mass function \( m(t, r) \) to the components of the energy-momentum tensor \( T^{\mu \nu} \) through a perturbative approach, where corrections appear linearly in terms of the accretion rate \( \dot{m} \). By introducing the back-reaction effect at the level of the EFEs, we maintain a first-order approximation that adequately captures the influence of accreting matter without violating the stability or consistency of the background metric.
	  
In this sense, our solution is self-consistent, as the derived mass function reflects only slight modifications in the BS horizon due to accretion, a scenario supported by the smallness of the perturbation terms \( m'(t, r) \) and \( \dot{m}(t, r) \). This is evident from Eq. \((\ref{012})\). Therefore, our solution is not only self-consistent but also corroborates the assumptions widely used in accretion theory, particularly in cylindrically symmetric spacetimes, where the energy-momentum tensor components remain compatible with the slowly varying behavior required by the underlying geometry.

	The energy conditions serve as fundamental constraints that encapsulate the general properties of most forms of matter, helping to eliminate many non-physical solutions to the EFEs. Among these conditions, we find the null, weak, dominant, and strong energy conditions \cite{r31}, all of which can be understood as limitations on the eigenvalues and eigenvectors associated with the energy-momentum tensor.
	 
{\bf	Null Energy Condition (NEC):} This condition states that for any future-pointing null vector field \( \vec{k} \), the energy density must satisfy \(\rho = T_{ab} k^a k^b \geq 0\).\\	 
{\bf Weak Energy Condition (WEC):} The WEC asserts that for every timelike vector field \( \vec{X} \), the energy density also meets the requirement \(\rho = T_{ab} X^a X^b \geq 0\).\\ 
{\bf Dominant Energy Condition (DEC):} In addition to fulfilling the weak energy condition for all future-pointing causal vector fields \( \vec{Y} \) (which can be either timelike or null), the DEC requires that the vector \( -T^b_a Y^b \) must remain a future-pointing causal vector.\\ 
{\bf Strong Energy Condition (SEC):} The SEC is defined such that for any timelike vector field \( \vec{X} \), the expression 
	 \begin{equation}
	 \left( T_{ab} - \frac{1}{2} T g_{ab} \right) X^a X^b \geq 0,
	\end{equation}
	 holds true.
These formulations do not depend on the specific matter source. However, if we consider an anisotropic fluid as the source, the energy-momentum tensor takes the form:
	\begin{equation}
	 T_{pq} = \left( \rho + p_{\perp} \right) u_p u_q + \left( p_{\|} - p_{\perp} \right) n_p n_q + p_{\perp} g_{pq},
    \end{equation}
	 where \( \rho \) denotes the energy density, \( u_q \) represents the four-velocity, and \( n_q \) is the spacelike unit vector. Here, \( p_{\|} \) and \( p_{\perp} \) correspond to the pressures parallel and perpendicular to \( n_q \), respectively.
	 For a static observer in this spacetime, the four-velocity will primarily point in the \( t \)-direction (time coordinate), and will have the following form,
	 \begin{equation}
	 u_q = \left( \frac{1}{\sqrt{-g_{tt}}}, 0, 0, 0 \right) = \left( \frac{1}{\sqrt{\alpha^{2}r^{2}-\frac{m(t,r)}{r}}}, 0, 0, 0 \right).
	 \end{equation}
	 The spacelike unit vector \( n_q \) will be orthogonal to \( u_q \), and you can choose it to point in the radial direction (in \( r \)) or along the angular direction \( \theta \). For simplicity, we consider \( n_q \) to be along the \( r \)-direction, yielding,
	 \begin{equation}
	 n_q = \left( 0, \frac{1}{\sqrt{g_{rr}}}, 0, 0 \right) = \left( 0, \sqrt{\alpha^{2}r^{2}-\frac{m(t,r)}{r}}, 0, 0 \right).
	 \end{equation} 
The relations \( u_q u^q = -1 \), \( n_q n^q = 1 \), and \( u_q n^q = 0 \) are essential conditions that must be fulfilled.

To check each of the energy conditions rigorously, we will use the components of the stress-energy tensor provided by Eqs. (\ref{07} - \ref{09}). We need to verify that these components satisfy the null, weak, strong, and dominant energy conditions.
 The NEC requires that for any null vector \( k^\mu \), we have,
\begin{equation}
T_{\mu \nu} k^\mu k^\nu \geq 0.
\end{equation}
Consider a null vector \( k^\mu = (k^0, k^1, 0, 0) \) with the condition \( g_{\mu \nu} k^\mu k^\nu = 0 \). Using Eqs. (\ref{07} - \ref{09}),
\begin{equation}
T_{\mu \nu} k^\mu k^\nu = \frac{1}{r^{2}} \big( 3\alpha^{2} r^{2} - m' \big) \left( (k^0)^2 + (k^1)^2 \right) + \frac{2 \dot{m}}{r^{2}} k^0 k^1.
\end{equation}
For the NEC to hold, \( 3\alpha^{2} r^{2} - m' \geq 0 \). So, the NEC is satisfied if,
\begin{equation}
 m' \leq 3\alpha^{2} r^{2}.
\end{equation}
The WEC requires \( T_{0}^{0} \geq 0 \). This implies \( \frac{1}{r^{2}} (3\alpha^{2} r^{2} - m') \geq 0 \), which leads to \( m' \leq 3\alpha^{2} r^{2} \), as in the NEC. For the timelike condition, for any timelike vector \( u^\mu \), the requirement is \( T_{\mu \nu} u^\mu u^\nu \geq 0 \). Using a vector  along the \( t \)-direction, \( u^\mu = (u^0, 0, 0, 0) \), with \( g_{\mu \nu} u^\mu u^\nu = -1 \), we find,
\begin{equation}
T_{\mu \nu} u^\mu u^\nu = T_{0}^{0} (u^0)^2= \frac{1}{r^{2}} (3\alpha^{2} r^{2} - m').
\end{equation}
Therefore, the WEC holds if \( m' \leq 3\alpha^{2} r^{2} \), which is consistent with the NEC.
The SEC is satisfied if,
\begin{equation}
(T_{\mu \nu} - \frac{1}{2} T g_{\mu \nu}) u^\mu u^\nu \geq 0,
\end{equation}
where, \( T = g^{\mu \nu} T_{\mu \nu} \) denotes the trace of the stress-energy tensor. Since \( T_{0}^{0} = T_{1}^{1} \) and the other terms depend on the mass function and its derivatives, these terms consistently reduce. Given the previous results, if \( m' \leq 3\alpha^{2} r^{2} \), the SEC is also satisfied. The DEC requires \( T_{0}^{0} \geq 0 \), which is already ensured by the WEC. The vector \( T^\mu_{\nu} u^\nu \) must be causal, and since the \( T_{0}^{1} \) term, representing flux, is compatible with the causal structure, the DEC holds under the same condition \( m' \leq 3\alpha^{2} r^{2} \). All energy conditions (NEC, WEC, SEC, and DEC) are satisfied for this metric if,
\begin{equation}
 m' \leq 3\alpha^{2} r^{2}.
\end{equation}
This inequality provides the constraint under which the energy conditions hold for our solution.
	 
\section{Analysis of Different Accerting Models}	
\subsection{Accretion of Dust}
In this section, we analyze the accretion of dust onto  BS, represented by the energy-momentum tensor of dust given by:
\begin{equation}
T_{j}^{i} = \rho V^{i} V_{j}, \label{017}
\end{equation}
where \(\rho\) is the rest mass density of the dust and \(V^i\) is the four-velocity of the dust particles. To ensure that the four-velocity is normalized, we impose the condition,
\begin{equation}
g_{ij} V^{i} V^{j} = -1, \label{018}
\end{equation}
indicating that \(V^i\) is a timelike vector. We define the four-velocity as:
\begin{equation}
V = \left[ \frac{1}{f_{0}}(\sqrt{u^{2} + f_{0}}), -u, 0, 0 \right], \label{019}
\end{equation}
where \(u\) represents the spatial component of the velocity, and \(f_0\) is a function characterizing the gravitational field of the BS. Using the normalization condition given by Eq. (\ref{018}) and the expression for the four-velocity given by Eq. (\ref{019}), we derive the components of the energy-momentum tensor as follows,
\begin{equation}
T_{0}^{0} = -\left(1 + \frac{u^{2}}{f_{0}}\right) \rho, \label{020}
\end{equation}
which represents the energy density of the dust, showing that it is affected by both the rest mass density \(\rho\) and the kinetic contribution from the dust's motion. For the momentum density, we have,
\begin{equation}
T_{0}^{1} = \rho V^{1} V_{0} = \rho u \sqrt{u^{2} + f_{0}}. \label{021}
\end{equation}
Eq. (\ref{021}) represents the flow of momentum due to the dust particles, indicating how their motion contributes to the overall momentum density in the spacetime. Substituting the expressions for \(T_{0}^{0}\) and \(T_{0}^{1}\) into the Eq. (\ref{013}) gives
\begin{align}
m(t,r) =& 8\pi r^{2} (r - r_{0}) \left(1 + \frac{u^{2}}{f_{0}}\right) \rho|_{r = r_{0}} + \alpha^{2} r^{3}\nonumber\\ 
&+ m_{0}+ 8\pi t r^{2} \rho u \sqrt{u^{2} + f_{0}}. \label{025}
\end{align}
The above equation can be interpreted as describing how the BS mass evolves with accreting dust. The first term represents the mass contribution from dust at a reference radius \( r_0 \) near the apparent horizon, influenced by its energy density. The initial mass is denoted by \( m_{0} \), while the last term accounts for the dynamic effect of the dust's momentum flow over time on the mass.

We present graphs of Eq. \ref{025},  the evolution of mass  as  time and a function of radial coordinates. The mass shows exponential growth over time while increasing linearly with the radial coordinate \( r \). This indicates that the central object is steadily accreting mass, with the rate of accretion intensifying as time progresses. The mass distribution remains relatively uniform along the radius, suggesting an isotropic or homogeneously distributed accretion flow, particularly along the radial axis. These results are depicted in Fig. 1(a) and (b).
\begin{figure*}[t]
\centering
\captionsetup[subfigure]{labelformat=empty} 
\begin{minipage}[t]{0.38\textwidth}
\centering
\includegraphics[width=1\linewidth]{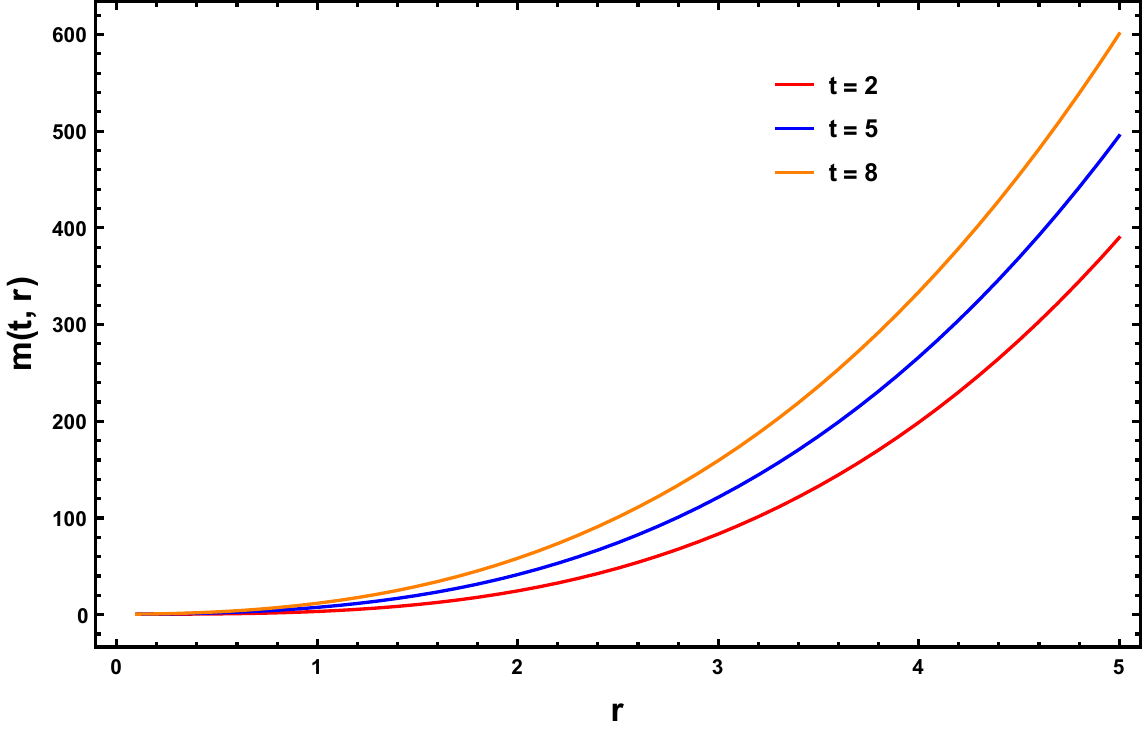}
\par\vspace{.3cm} Fig. 1(a)
\end{minipage}
\hspace{1.8cm}
\begin{minipage}[t]{0.38\textwidth}
\centering
\includegraphics[width=1\linewidth]{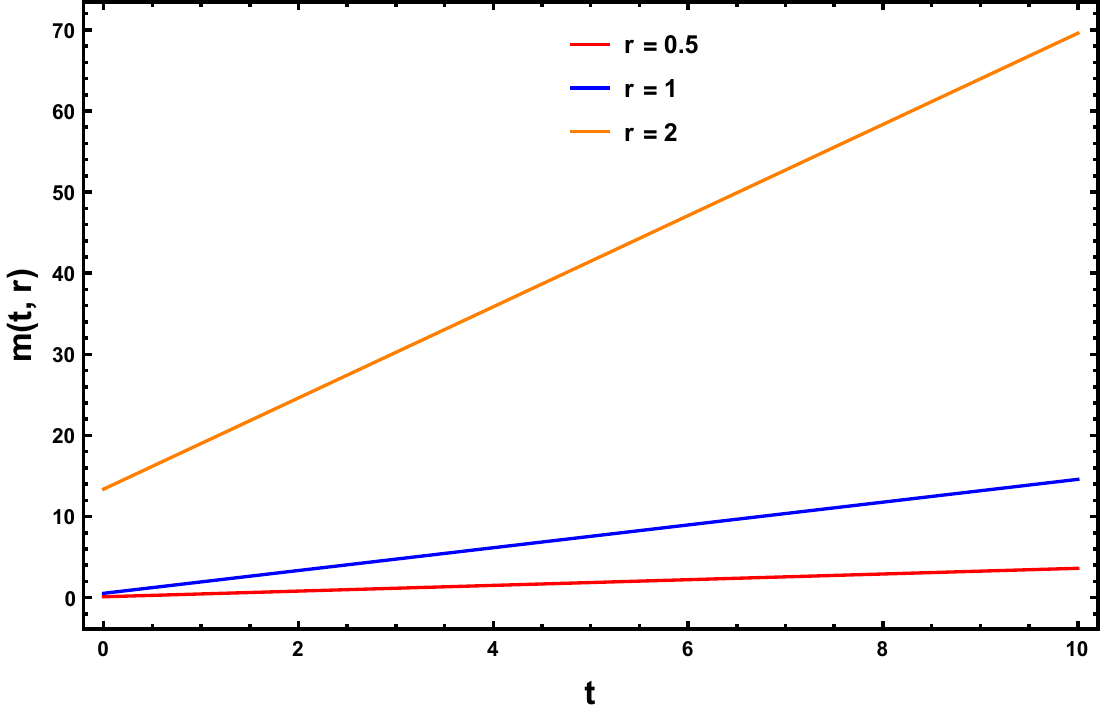}
\par\vspace{.3cm} Fig. 1(b)
\end{minipage}%
\vspace{.7cm}
\begin{minipage}[t]{0.38\textwidth}
\centering
\includegraphics[width=1\linewidth]{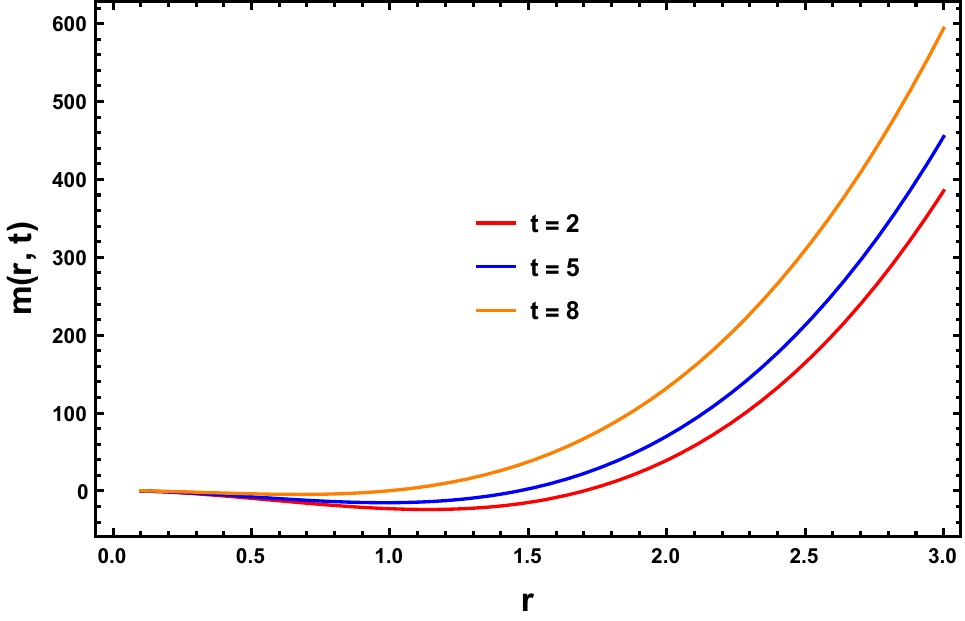}
\par\vspace{.3cm} Fig. 1(c)
\end{minipage}%
\hspace{1.8cm}
\begin{minipage}[t]{0.38\textwidth}
\centering
\includegraphics[width=1\linewidth]{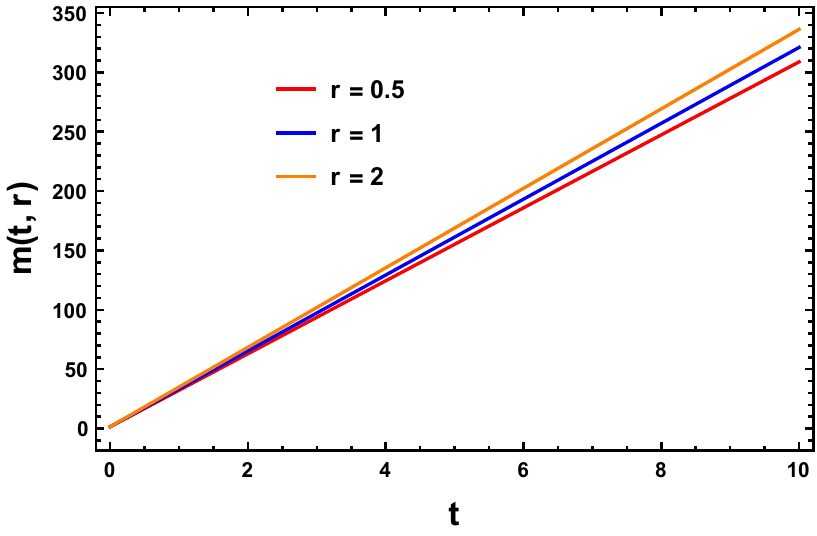}
\par\vspace{.3cm} Fig. 1(d)
\end{minipage}%
\vspace{.7cm}
\begin{minipage}[t]{0.38\textwidth}
\centering
\includegraphics[width=1\linewidth]{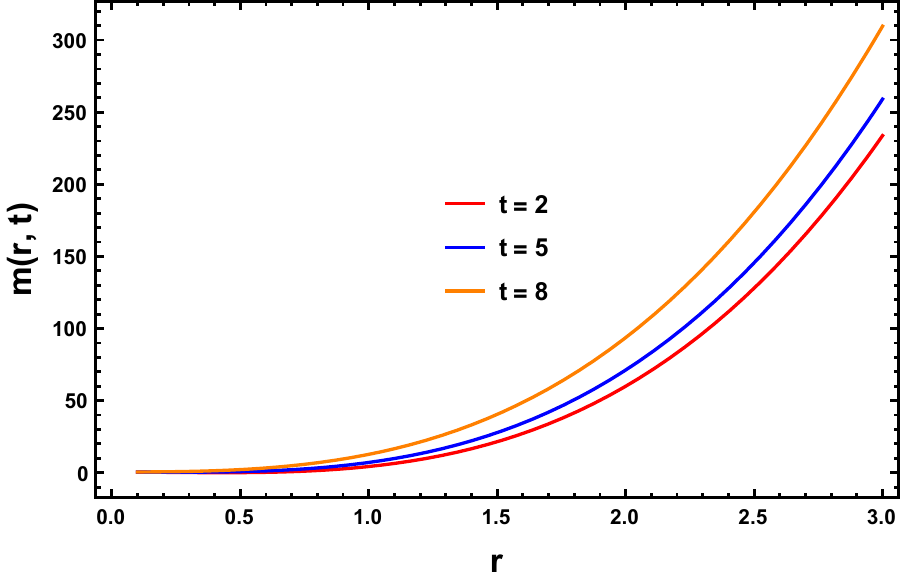}
\par\vspace{.3cm} Fig. 1(e)
\end{minipage}%
\hspace{1.8cm}
\begin{minipage}[t]{0.38\textwidth}
\centering
\includegraphics[width=1\linewidth]{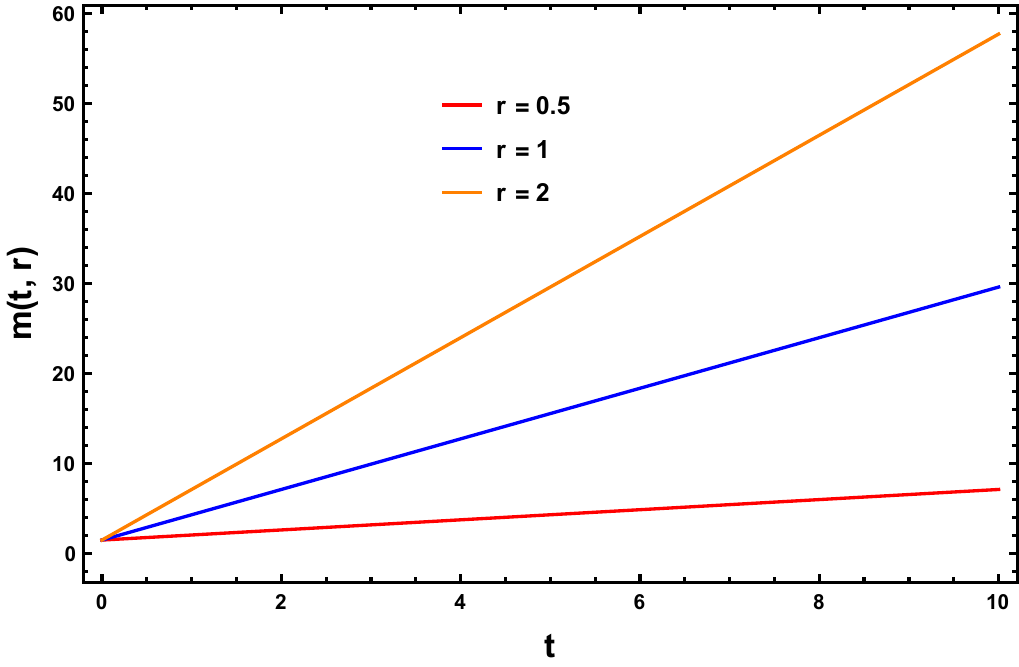}
\par\vspace{.3cm} Fig. 1(f)
\end{minipage}
	\captionsetup{justification=raggedright,singlelinecheck=false}
\caption{ The corrected mass function \( m(t, r) \) is plotted against time \( t \) for various fixed radial coordinates, illustrating different accretion behaviors across models. In Fig. 1(a), \( m(t, r) \) is shown for different radial values , where mass accretion increases with larger \( r \). Fig. 1(b) demonstrates a nearly linear increase in \( m(t, r) \) over time, characteristic of a dust model, with higher accretion rates at larger radii. In Fig. 1(c), a non-linear growth in \( m(t, r) \) suggests additional pressure effects within a perfect fluid model, impacting the accretion rate. Fig. 1(d) presents \( m(t, r) \) at small radial values, where all models show similar behavior, likely due to dominant gravitational effects near the center. Finally, Fig. 1(e) shows the mass function over time for larger radii, where the radiative fluid model leads to accelerated accretion in outer regions due to radiation pressure.}
\end{figure*}

\subsection{Accretion of Perfect Fluid}
The energy-momentum tensor for a perfect fluid, which characterizes the distribution of energy and momentum in a fluid at rest or in motion, is expressed as,
\begin{equation}
T_{j}^{i} = (\rho + P)V^{i}V_{j} + P\delta_{j}^{i}, \label{026}
\end{equation}
where \(\rho\) is the energy density, \(P\) is the pressure, \(V^{i}\) is the fluid's 4-velocity, and \(\delta_{j}^{i}\) is the Kronecker delta. Using the Eqs. (\ref{019}) and (\ref{026}), we obtain the following specific components of energy momentum tensor,
\begin{equation}
T_{0}^{0} = -\left(1 + \frac{u^{2}}{f_{0}}\right)\rho - \frac{u^{2}}{f_{0}}P, \label{027}
\end{equation}
\begin{equation}
T_{0}^{1} = (\rho + P)u\sqrt{u^{2} + f_{0}}.
\label{028}
\end{equation}
These components reveal how the energy density and pressure of the fluid contribute to the overall energy-momentum distribution in a dynamic setting, particularly under the influence of velocity \(u\) and the gravitational field characterized by \(f_{0}\). Substituting Eqs (\ref{027}) and (\ref{028}) into the mass function given by Eq. (\ref{013}), we derive the corrected mass  for perfect fluid case as follows,
\begin{align}
m(t, r) = & \ 8\pi r^{2}(r - r_{0})\left[\left(1 + \frac{u^{2}}{f_{0}}\right)\rho + \frac{u^{2}}{f_{0}}P\right]_{r = r_{0}} + \alpha^{2}r^{3} \nonumber \\
& + m_{0} + 8\pi tr^{2}(\rho + P)u\sqrt{u^{2} + f_{0}}. \tag{7}
\end{align}
This equation for \(m(t, r)\) shows the relation between the fluids dynamic properties and the gravitational field. The terms involving \(\rho\) and \(P\) indicate that the mass perceived by an observer can be significantly influenced by the fluid's velocity and pressure, revealing the rich structure of gravitational interactions in a perfect fluid context. Furthermore, the dependence on \(r\) suggests that mass accumulation is affected not just by the local energy density but also by the spatial distribution of the fluid and its motion relative to the observer.

Fig. 1(c) and 1(d) illustrate the variation of the corrected mass as a function of both coordinates, \( t \) and \( r \). The graphs show that the corrected mass increases with respect to both \( t \) and \( r \). Initially, as \( t \) increases, the mass grows rapidly, indicating a strong dependency on the temporal coordinate. For smaller values of \( t \) close to zero, the mass still increases, though at a slower rate.  Similarly, with respect to \( r \), the corrected mass initially grows linearly, reflecting a direct proportionality. As \( r \) increases further, the various mass profiles begin to converge toward a single line, indicating that the radial dependency becomes less significant at larger radii. At smaller \( r \) values, the mass increase is more noticeable, with the slope of each line changing accordingly, suggesting a dynamic relationship between mass accumulation and radial distance in this model.

\subsection{Accretion of Radiative Fluid}

In the case of a radiative fluid, the equation of state is given by \(\rho = 3P\), which characterizes it as a perfect fluid where the energy density \(\rho\) is proportional to three times the pressure \(P\). For such a fluid, the energy-momentum tensor takes the form,
\begin{equation}
T^i_j = 4P V^i V_j + P \delta^i_j.
\end{equation}
 Using the similar procedure, we find the components of the energy-momentum tensor,
\begin{equation}
T^0_0 = -3P - \frac{4u^2}{f_0}P,~~T^0_1 = 4P u \sqrt{u^2 + f_0},
\end{equation}
where \(u\) represents the radial velocity component, and \(f_0\) is a function associated with the metric components, typically representing gravitational effects on the fluid flow. Substituting these components into Eq. (\ref{013}) yields the corrected mass function, which incorporates the effects of both radial motion and gravitational interactions given by
\begin{align}
m(t, r) =& 8 \pi r^2 (r - r_0)(3P + \frac{4u^2}{f_0} P) \bigg|_{r = r_0} + \alpha^2 r^3 \\\nonumber
&+ m_0 + 8 \pi t r^2 \cdot 4P u \sqrt{u^2 + f_0}.
\label{032}
\end{align}
In this case, the corrected mass exhibits a similar behavior to that in previous cases. The mass increases with both \(r\) and \(t\): initially, it rises linearly with \(r\) up to \(r=1\), after which it grows exponentially for \(r>1\). Along the time coordinate, the mass shows a linear increase for all values of \(t\). This behavior can be seen in Fig. 1(e) and (f).
\subsection{Corrected Apparent Horizon}
The position of the apparent horizon depends on the choice of a coordinate system. For the metric (\ref{3}), it can be shown that the location of the
 apparent horizon, $r_{h}$, can be found as
\begin{equation}
r_h \approx \left( \frac{ m(t, r_h)}{\alpha^2} \right)^{\frac{1}{3}}.
\end{equation} 
Using the value of corrected mass we have,
\begin{equation}
r_h \approx \left( \frac{-8\pi r_h^{2}(r_h - r_{0}) T_{0}^{0} + m_{0}}{\alpha^2} + \frac{8\pi t r_h^{2} T_{0}^{1}}{\alpha^2} + r_h^{3} \right)^{\frac{1}{3}},
\end{equation}
further simplifying the above equation using the binomial approximation, we get the expression for corrected apparent horizon for the line element  (\ref{013}),
\begin{equation}
r_h \approx r_h + \frac{1}{3 \alpha^2 r_h^{2}} \left( -8\pi r_h^{2}(r_h - r_{0}) T_{0}^{0} + m_{0} + 8\pi t r_h^{2} T_{0}^{1} \right).
\end{equation}

We presented 3D plot, where the corrected apparent horizon, \( r_h \), is depicted in terms of time \( t \) and either the energy density \( \rho \) or pressure \( P \) for three distinct fluid models relevant to  accretion. Fig. 3(a) specifically illustrates the relationship between \( r_h \), \( t \), and \( \rho \) for a dust model, where dust is characterized by a pressureless medium (\( P = 0 \)). In this scenario, the horizon gradually shifts over time, indicating a steady, unresisted accretion process due to the absence of internal pressure in the dust. As commonly discussed in the literature \cite{r16,r24}, dust models serve as simplified representations of accretion, where the lack of opposing pressure allows the BH to gain mass and energy smoothly, leading to a consistent, linear expansion of the corrected horizon as the accretion continues over time.

Figs. 3(b) and 3(c)  show the corrected horizon \( r_h \) with respect to time \( t \) and pressure \( P \) or \( \rho \) in a perfect fluid model, which has both energy density and pressure (non-zero \( P \)). The presence of pressure introduces resistance in the accretion process, affecting the evolution of the event horizon. In GR, the presence of pressure leads to a more complex interaction with the BH's gravitational field. Consequently, as shown in the graphs, \( r_h \) exhibits a steeper growth with time when compared to the dust model, due to the additional energy-momentum components associated with pressure. As a source of dynamic back-reaction, where the pressure of the fluid influences the BS corrected apparent horizon. For instance, for higher pressures, the back-reaction effect can slow down the rate of accretion due to repulsive forces that counteract the BS gravitational pull.

Fig. 3(d) provides the case of the radiative fluid here, \( r_h \) is plotted with respect to \( t \) and \( P \). Radiative fluids, often represented by radiation pressure, involve particles moving at nearly the speed of light. This high-speed inflow creates a different profile for the corrected apparent horizon's growth. The accretion rate can vary significantly because radiation exerts strong pressure, potentially heating the surrounding medium and affecting the accretion dynamics. Intuitively, radiation pressure can significantly limit or even reverse the in fall of matter under certain conditions. In this model, \( r_h \) shows a steep slope, suggesting that the horizon size could either increase rapidly or stabilize depending on the balance between gravitational attraction and radiation pressure.

The relationship between the apparent horizons and the event horizon, as depicted in Fig. 2, reveals their distinct but interrelated roles in the spacetime structure of the BS. The event horizon, shown in blue, serves as the ultimate causal boundary, beyond which no signals or matter can escape to infinity.  It marks the point of no return, where causal connections with the external universe are severed. As time progresses, the event horizon moves inward, reflecting the increasing collapse of the BS spacetime. The apparent horizons (green dashed, orange dot-dashed, and red dotted lines for dust, radiative fluid, and perfect fluid, respectively) represent the boundaries where light cones collapse, signaling the regions within the BS where the spacetime curvature becomes extreme. These apparent horizons are not static; they evolve over time, moving closer to the event horizon. The movement of these horizons reflects how different types of matter (dust, radiative fluid, and perfect fluid) interact with the BS gravitational field and influence the spacetime structure.

From a physical perspective, the dust case, represented by the green dashed line, reflects a more gradual and weak gravitational collapse due to the absence of pressure in dust matter. The apparent horizon in this case moves inward slowly, as the dust contributes minimally to the curvature of spacetime. In contrast, the radiative fluid (orange dot-dashed line) exhibits stronger gravitational effects due to the pressure and energy flux associated with radiation, causing the apparent horizon to contract more rapidly. The perfect fluid (red dotted line), which has both energy density and pressure, produces the most intense gravitational influence, leading to the most rapid contraction of the apparent horizon. In all three cases, the apparent horizons approach the event horizon over time, but they do not coincide, suggesting that the event horizon remains the definitive boundary of the BS, while the apparent horizons provide a time-dependent, matter-specific description of the BS internal structure. This demonstrates the intricate relationship between matter distribution, spacetime curvature, and the dynamic evolution of BS horizons  highlights how matter accretion affects the trapped region, with the apparent horizon adapting to local energy conditions while the event horizon remains a global, coordinate-independent feature of the BS geometry.
\begin{figure}
\includegraphics[width=1\linewidth]{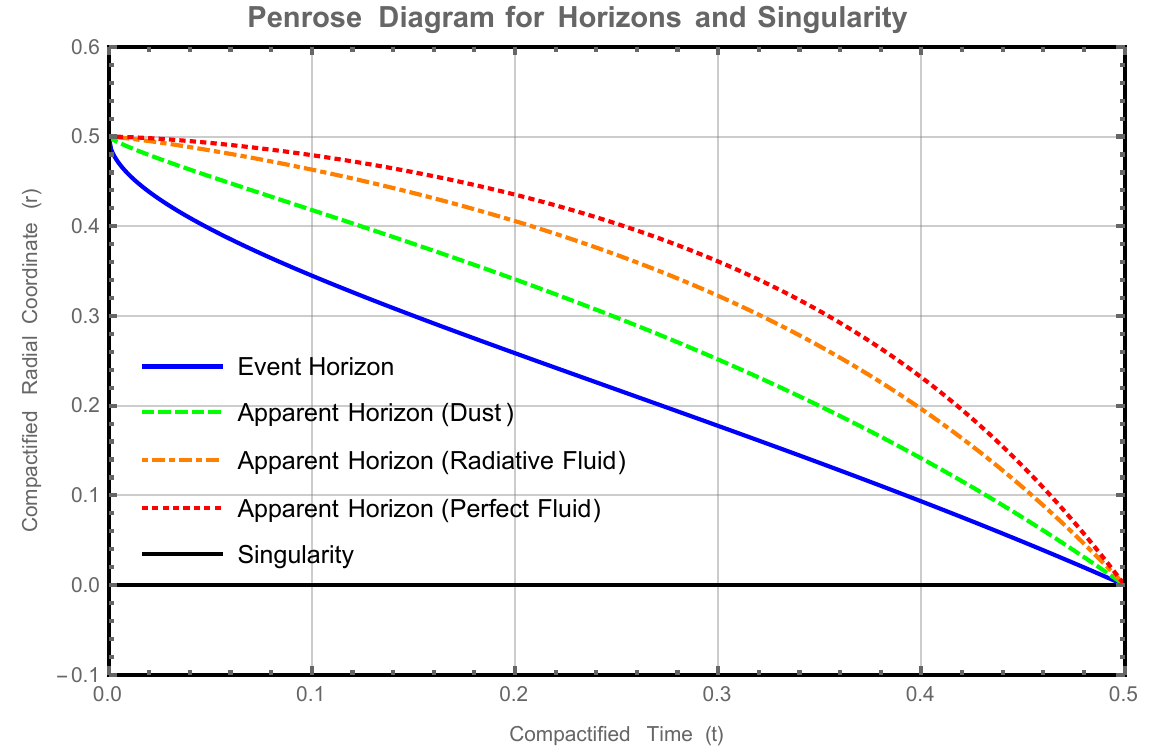}
\captionsetup{justification=raggedright,singlelinecheck=false}
\caption{The Penrose diagram showing the evolution of event horizon and apparent horizons for different types of matter: dust (green dashed line), radiative fluid (orange dot-dashed line), and perfect fluid (red dotted line). The event horizon is depicted as a blue curve, marking the boundary beyond which nothing can escape the BS gravitational pull. The apparent horizons for each type of matter move inward over time, with the perfect fluid causing the most rapid contraction. The singularity, marked by a black line at \(r = 0\), represents the final collapse point at the center of the BS. The diagram provides the dynamic relationship between the event horizon, apparent horizons, and the singularity in the context of the BS spacetime.}
\end{figure}

\begin{figure*}[t]
\centering
\begin{minipage}[t]{0.38\textwidth}
\centering
\includegraphics[width=1\linewidth]{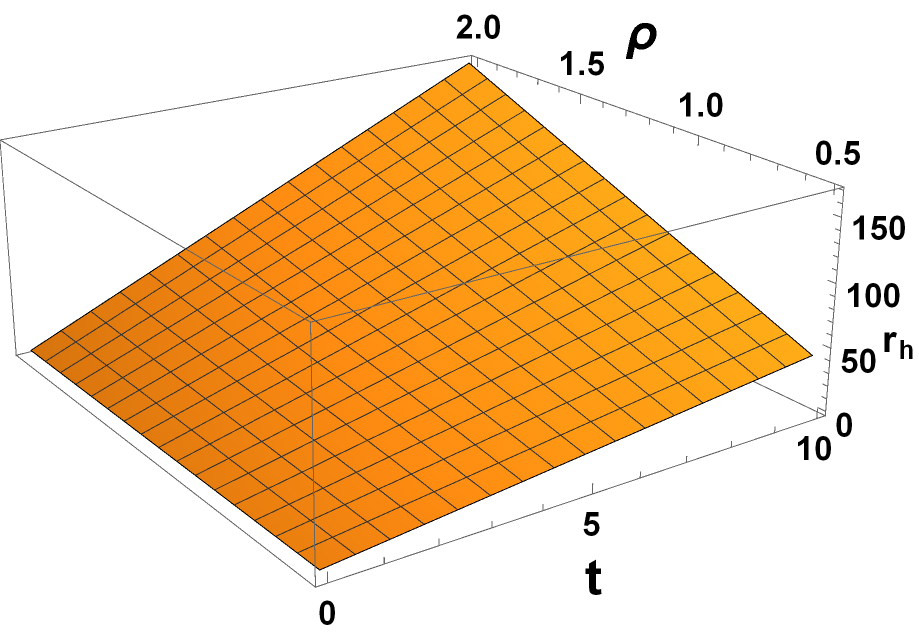}
\caption*{Fig. 3(a)}
\end{minipage}
\hspace{1.8cm}
\begin{minipage}[t]{0.38\textwidth}
\centering
\includegraphics[width=1\linewidth]{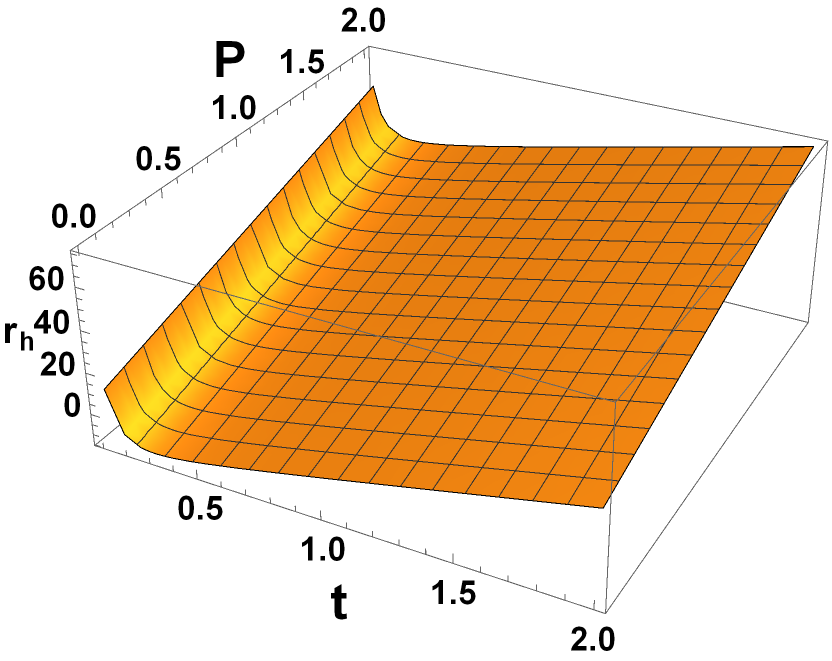}
\caption*{Fig. 3(b)}
\end{minipage}%
	 		
	 		~
\vspace{.7cm}
\begin{minipage}[t]{0.38\textwidth}
\centering
\includegraphics[width=1\linewidth]{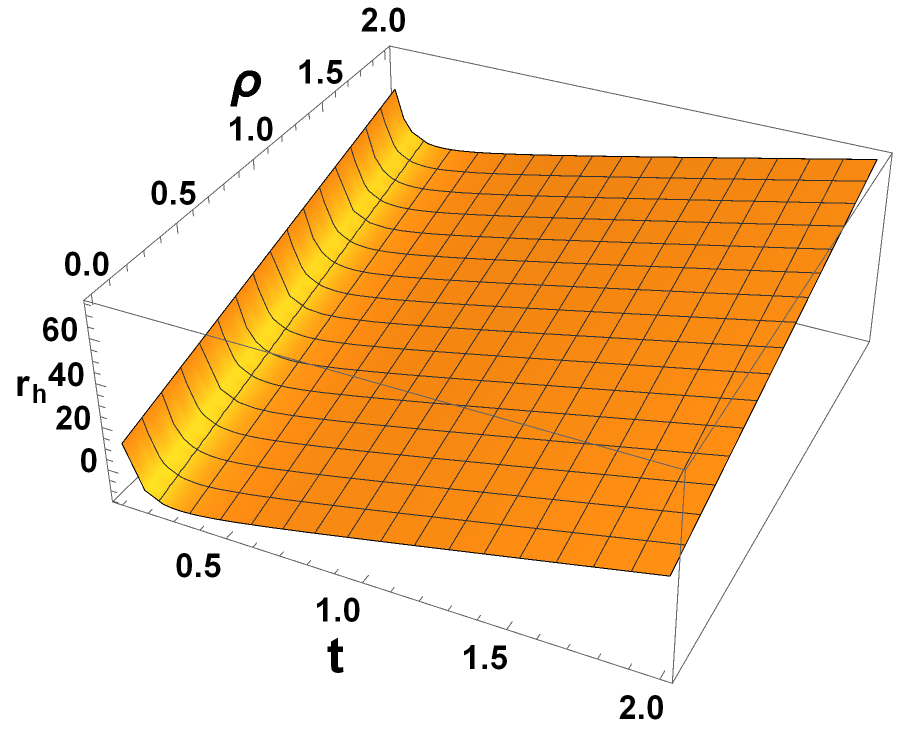}
\caption*{Fig. 3(c)}
\end{minipage}%
\hspace{1.8cm}
\begin{minipage}[t]{0.38\textwidth}
\centering
\includegraphics[width=1\linewidth]{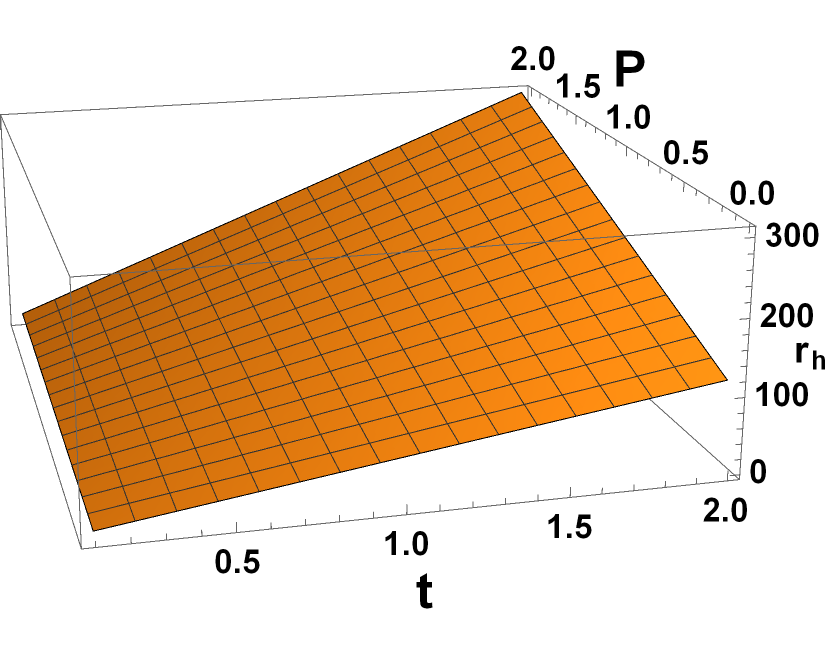}
\caption*{Fig. 3(d)}
\end{minipage}%
\\	~
	\captionsetup{justification=raggedright,singlelinecheck=false}
\caption{ The corrected apparent horizon \( r_h \) plotted against time \( t \), energy density \( \rho \), and pressure \( P \) for different accretion models, each providing distinct insights into BS growth. Fig. 3(a) depicts the dust model, representing the simplest case with a steady accretion rate due to the absence of pressure. Fig. 3(b) and Fig. 3(c) display the perfect fluid model, where pressure moderates accretion, leading to back-reaction effects on the system. Finally, Fig. 3(d) shows the radiative fluid model, where high radiation pressure induces complex interactions, significantly affecting accretion rates and stability through energy transfer and radiation dynamics.}
\end{figure*}	
\subsection{Corrected Entropy}
To find the expression for entropy \(S\), we use the Bekenstein-Hawking entropy formula \cite{r33}, which is given by
\begin{equation}
S = \frac{k_B A}{4 \, l_p^2},
\end{equation}
where \(A\) is the area of the BS horizon, \(k_B\) is Boltzmann’s constant, and \(l_p\) is the Planck length. For simplicity, we often use units where \(k_B = 1\) and \(l_p = 1\), so the entropy is directly proportional to the horizon area as
\begin{equation}
S = \frac{A}{4},\label{036}
\end{equation}
ands the area of the event horizon can find as, 
\begin{equation}
A = \int_0^{2\pi} d\theta \int_{z_1}^{z_2} dz \, (\alpha r^2).\label{037}
\end{equation}

If we assume the BS extends infinitely along the \(z\)-direction, the integration over \(z\) would contribute a factor proportional to the length of the \(z\)-direction, which is infinite. This length is denoted as \(L_z\). Thus, the infinite extent of the \(z\)-direction naturally introduces \(L_z\) as a scaling factor in the calculations. Specifically, the line element \(\ref{3}\) describes a BS with a cylindrical horizon, assuming the coordinate \(z\) spans the entire real line, i.e., \( -\infty < z < \infty \). However, if the \(z\)-coordinate is restricted to the interval \(0 \leq z < 2\pi\), the configuration corresponds to a closed BS with a toroidal horizon, where the \(z\)-coordinate is periodic, and the horizon takes the shape of a torus \cite{r32}. Therefore we get,
\begin{equation}
A = (2\pi) \cdot (\alpha r_h^2) \cdot L_z.\label{038}
\end{equation}
Using Eqs. (\ref{037}) and (\ref{038}) in(\ref{036}), we have
\begin{equation}
S = \frac{A}{4} = \frac{1}{4} \left( 2\pi \alpha r_h^2 L_z \right) = \frac{\pi \alpha r_h^2 L_z}{2}.
\label{0040}
\end{equation}
\(L_z\) is the length in the \(z\) direction, which could be treated as a constant depending on the physical scenario. 

In Fig. 4(a), entropy is plotted as a function of time \(t\) and density \(\rho\) for dust. The surface illustrates an increase in entropy as both time and density rise. This suggests that, over time, accumulating dust leads to higher density, which contributes to gravitational potential and, consequently, raises entropy. This outcome aligns with the second law of thermodynamics, which states that entropy in an isolated system generally increases over time due to the increasing matter density.

Fig. 4(b) and Fig. 4(c) display corrected entropy for the perfect fluid model, shown as a function of in Fig. 4(b) and as a function of pressure \(P\) and time \(t\) in Fig. 4(c). Fig. 4(b) exhibits a similar trend to the dust model, with entropy rising over time and with density, reflecting the added effects of pressure within the fluid. This pressure contributes to gravitational potential and energy density, thereby enhancing entropy generation. 

In Fig. 4(c), entropy’s relationship with pressure \(P\) and time \(t\) reveals that higher pressures correspond to greater entropy increases. This pattern highlights pressure's role in energy exchanges within the fluid, enhancing entropy through compressibility and fluid interactions, thus amplifying entropy production in perfect fluids. Fig. 4(d) presents corrected entropy for the radiative fluid model, shown as a function of pressure \(P\) and time \(t\). Here, entropy increases sharply with both pressure and time, underscoring the significant influence of radiation. Radiative fluids dissipate energy through radiation, leading to a rapid entropy increase. The steep gradient indicates that entropy in radiative fluids is highly responsive to changes in pressure and time, showing a more pronounced rise compared to the dust and perfect fluid models.
\begin{figure*}[t]
	\centering
	\begin{minipage}[t]{0.38\textwidth}
		\centering
		\includegraphics[width=1\linewidth]{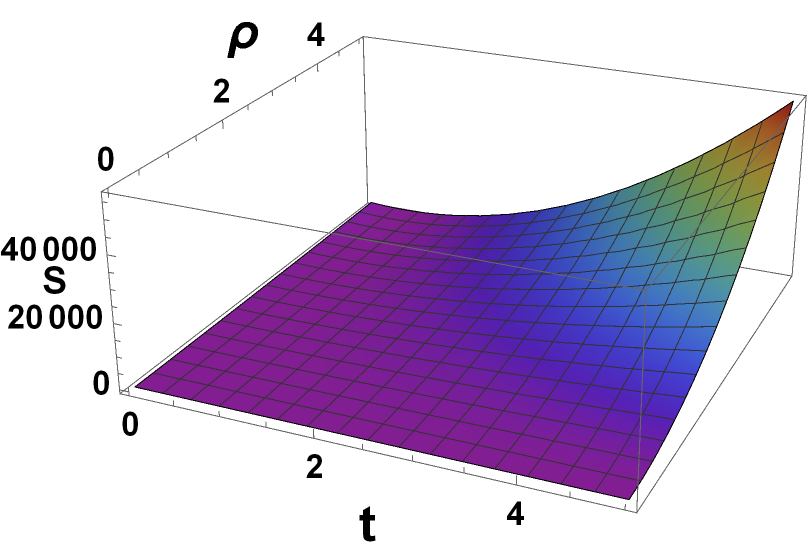}
		\caption*{Fig. 4(a)}
	\end{minipage}
	\hspace{1.8cm}
	\begin{minipage}[t]{0.38\textwidth}
		\centering
		\includegraphics[width=1\linewidth]{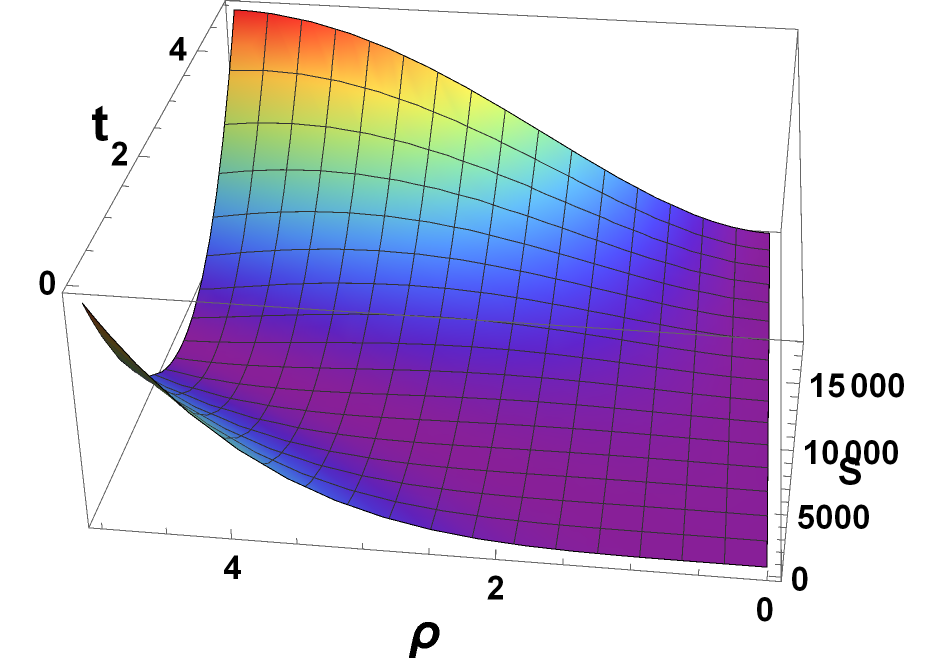}
		\caption*{Fig. 4(b)}
	\end{minipage}%
	
	~
	\vspace{.7cm}
	\begin{minipage}[t]{0.38\textwidth}
		\centering
		\includegraphics[width=1\linewidth]{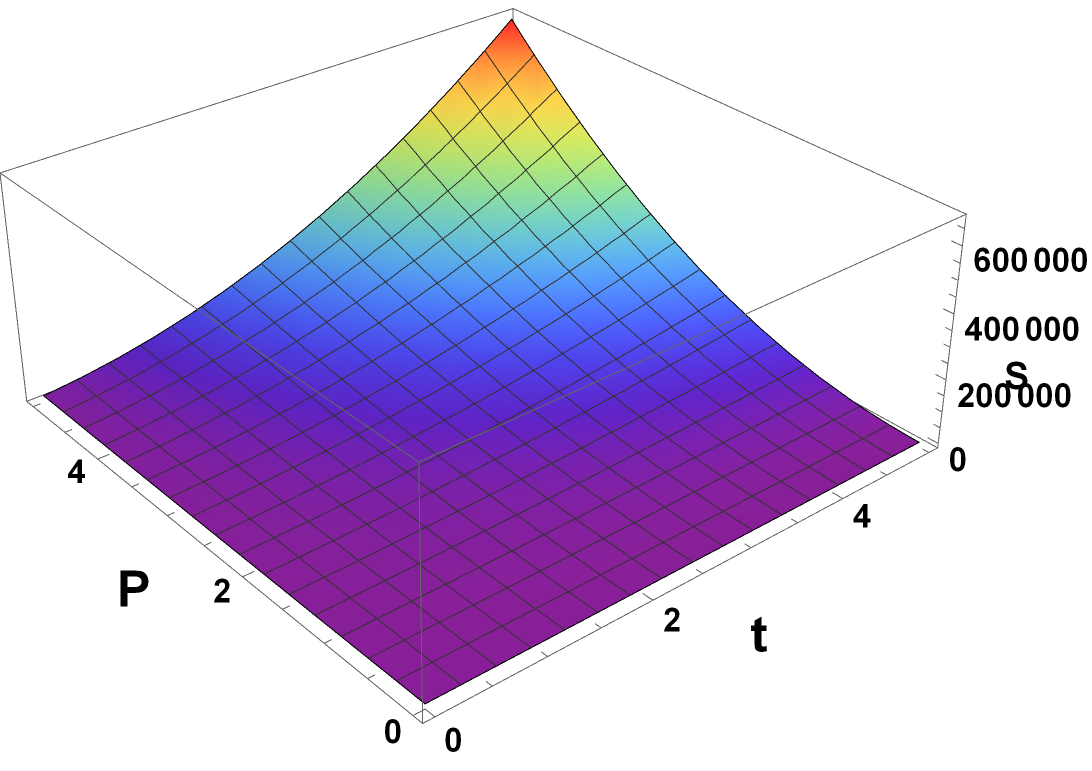}
		\caption*{Fig. 4(c)}
	\end{minipage}%
	\hspace{1.8cm}
	\begin{minipage}[t]{0.38\textwidth}
		\centering
		\includegraphics[width=1\linewidth]{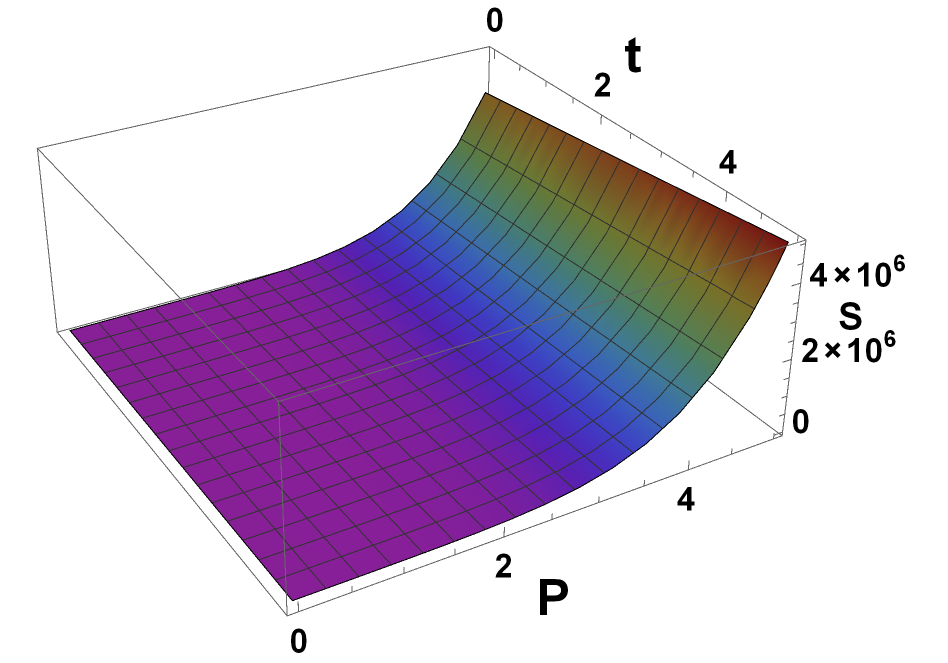}
		\caption*{Fig. 4(d)}
	\end{minipage}%
	\\	~
		\captionsetup{justification=raggedright,singlelinecheck=false}
	\caption{The corrected entropy \( S \) as a function of time \( t \), energy density \( \rho \), and pressure \( P \) for various accretion models, highlighting thermodynamic principles in BS growth. Fig. 4(a) illustrates the dust model, where entropy increases primarily due to rising density, with minimal internal interactions. Fig. 4(b) and Fig. 4(c) display the perfect fluid model, where pressure contributes to entropy production by enhancing internal energy exchanges, resulting in greater entropy as pressure and density increase. Fig. 4(d) depicts the radiative fluid model, showing rapid entropy growth due to radiative energy dissipation, which intensifies entropy  with increasing pressure. These trends reveal how different matter types and energy interactions influence entropy evolution, providing insights into accretion dynamics and back-reaction effects in astrophysical contexts.}
\end{figure*}

\subsection{Corrected Temperature}

To calculate the temperature, we will usey the concept of Hawking radiation, particularly relevant for BHs. The temperature can be derived from the metric's geometry, specifically focusing on the surface gravity of the event horizon. The surface gravity \(\kappa\) at the event horizon is given by,
\begin{equation}
\kappa = -\frac{1}{2} \left. \frac{dg_{tt}}{dr} \right|_{r=r_h}.
\label{040}
\end{equation}
Evaluating at the horizon \( r = r_h \), where \( g_{tt} = 0 \), we obtain,
\begin{equation}
\kappa = -\frac{1}{2} \left(-\alpha^{2}(2r_h - 3)\right) = \frac{\alpha^{2}(2r_h - 3)}{2}.
\end{equation}
The corresponding temperature \( T \) is then
\begin{equation}
T = \frac{\kappa}{2\pi}.
\label{042}
\end{equation}
Thus, the temperature \( T \) is basically  related to the location of the correct apparent horizon \( r_h \), 
\begin{equation}
T = \frac{\alpha^{2}(2r_h - 3)}{4\pi}.
\end{equation}
The temperature \( T \) of the corrected apparent horizon \( r_h \) for three different accretion models—dust, perfect fluid, and radiative fluid—is discussed graphically. In BH thermodynamics, the temperature of the horizon is typically inversely related to the horizon radius, so as \( r_h \) increases, \( T \) tends to decrease.

Fig. 5(a) shows the temperature for the dust fluid model, where the temperature \( T \) of the corrected apparent horizon drops sharply as \( r_h \) increases, eventually stabilizing at a low value. With dust lacking pressure, the accretion process is straightforward, leading to a rapid expansion of \( r_h \) and a corresponding decline in \( T \). This trend reflects the cooling effect due to the growing horizon, with temperature inversely scaling with horizon radius. In Fig. 5(b), for the perfect fluid model, a similar pattern is observed as \( T \) decreases with increasing \( r_h \), though the gradient is slightly different from the dust model. Here, the presence of pressure moderates the rate of accretion, causing a more gradual increase in \( r_h \). This moderation results in a smoother cooling effect, where the temperature decline is less abrupt. The pressure in a perfect fluid acts as a resistive force, influencing the BS thermal behavior by limiting rapid horizon expansion.

Finally, Fig. 5(c) shows the radiative fluid model. In this case, the temperature \( T \) also decreases with \( r_h \), but the curve differs significantly from the previous models, showing a slower decline at larger \( r_h \) values. The high-energy particles and radiation pressure in radiative fluids create a unique thermodynamic balance, affecting the black hole's properties in a distinct way. This slower temperature decrease aligns with findings in the literature \cite{r24, r27}, where radiative fluids are often associated with steady-state horizon expansion due to radiation’s outward pressure, which slows horizon growth and maintains a higher temperature for an extended period compared to other models.
	\begin{figure*}[t]
		\centering
		\begin{minipage}[t]{0.38\textwidth}
			\centering
			\includegraphics[width=1\linewidth]{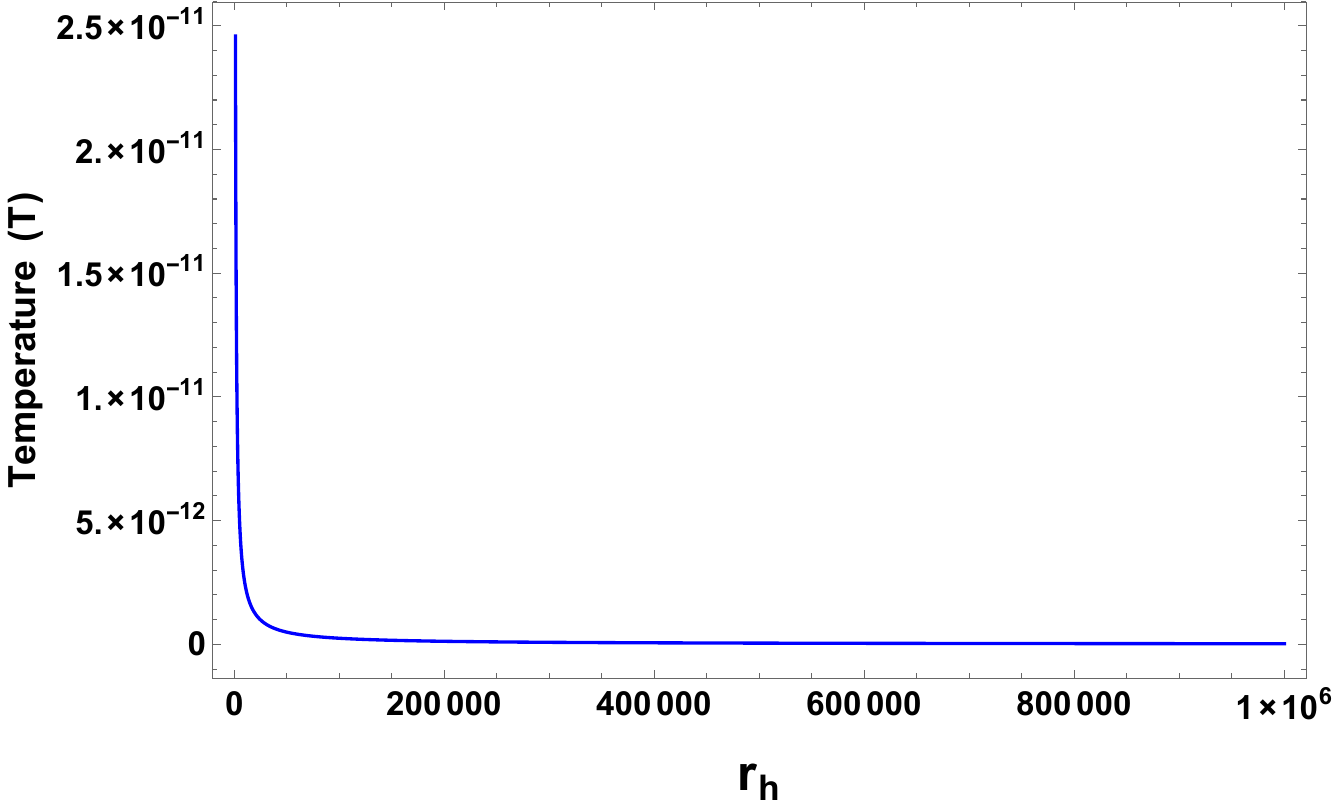}
			\caption*{Fig. 5(a)}
		\end{minipage}
		\hspace{1.8cm}
		\begin{minipage}[t]{0.38\textwidth}
			\centering
			\includegraphics[width=1\linewidth]{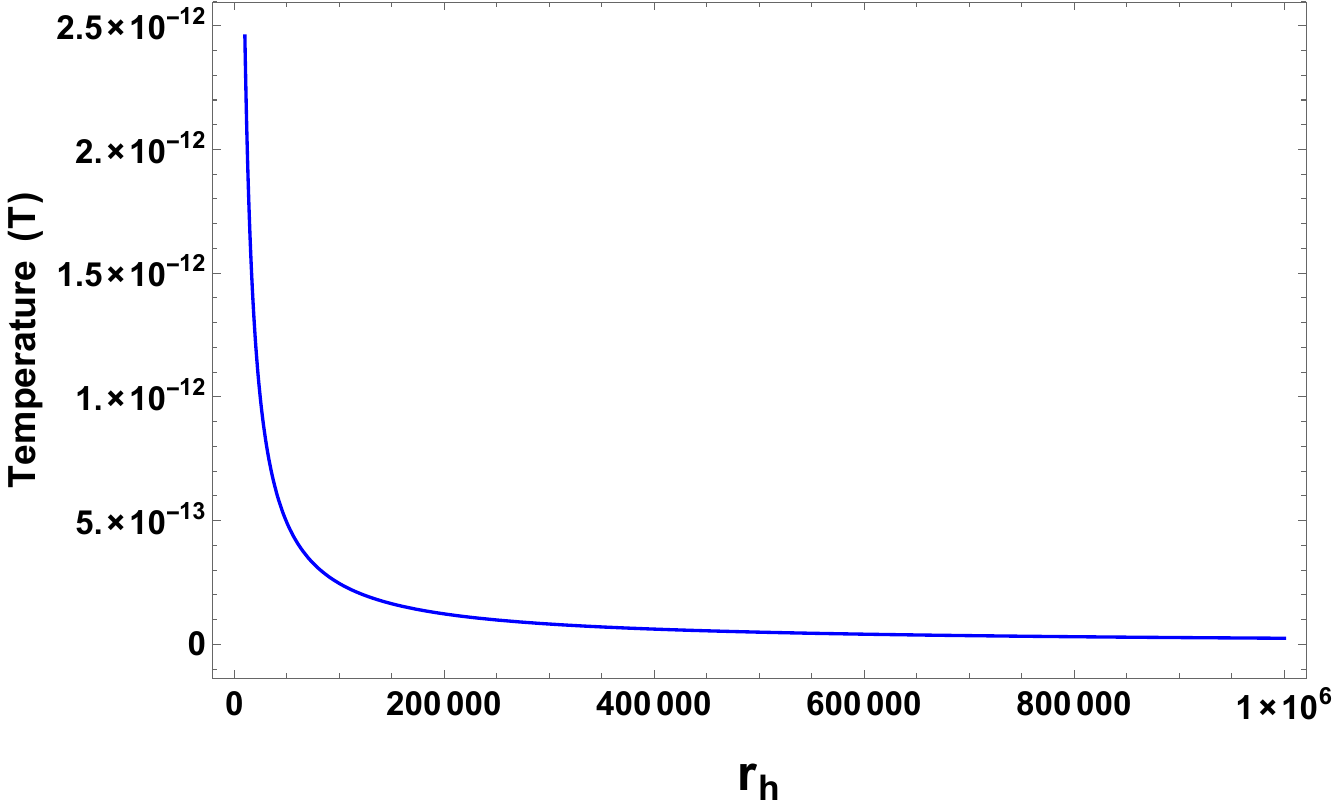}
			\caption*{Fig. 5(b)}
		\end{minipage}%
		~
		
		\vspace{.7cm}
		\begin{minipage}[t]{0.38\textwidth}
			\centering
			\includegraphics[width=1\linewidth]{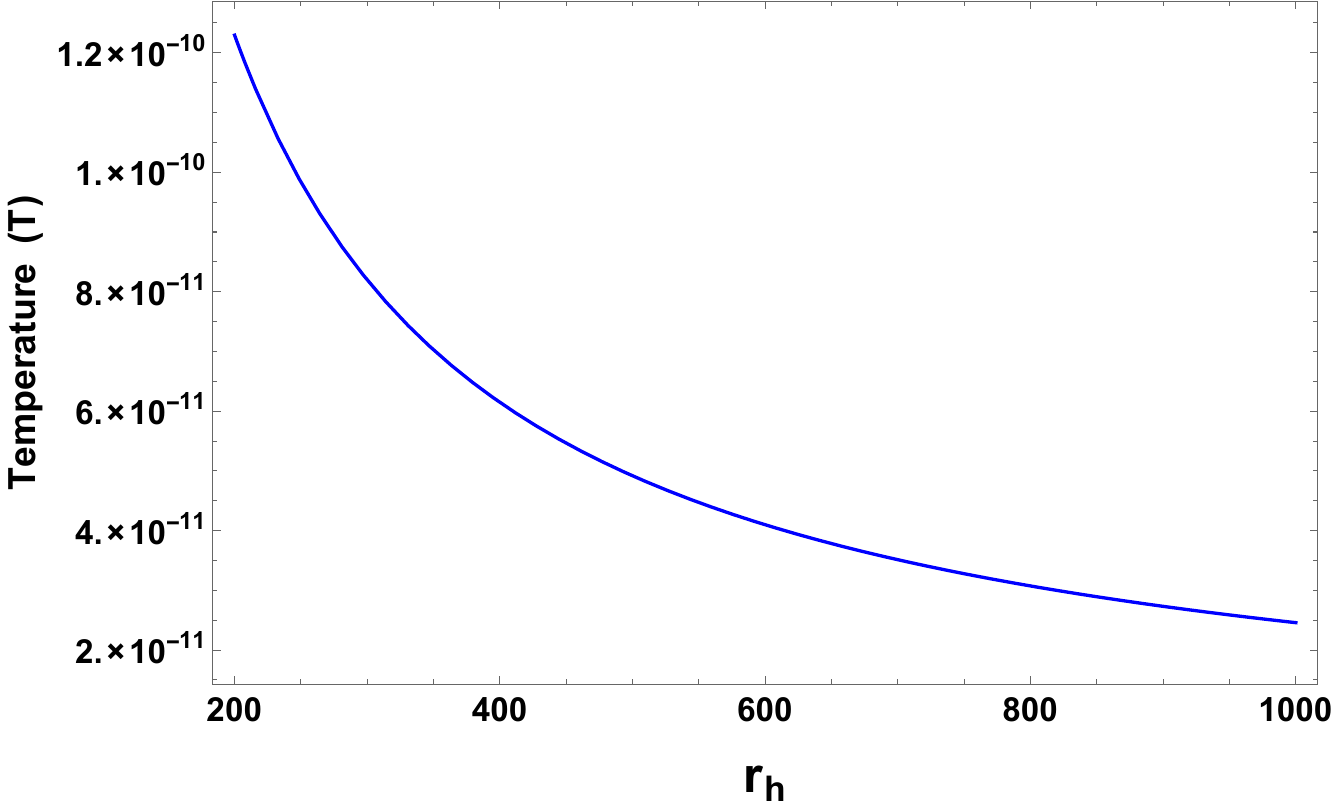}
			\caption*{Fig. 5(c)}
		\end{minipage}%
		\captionsetup{justification=raggedright,singlelinecheck=false}
	\caption{ The corrected temperature \( T \) as a function of the apparent horizon \( r_h \) for different accretion models, showing how each model impacts temperature evolution. Fig. 5(a) illustrates the dust model, where rapid horizon growth leads to a quick temperature drop. Fig. 5(b) presents the perfect fluid model, in which pressure moderates horizon expansion, resulting in a more gradual temperature decrease. Fig. 5(c) shows the radiative fluid model, where radiation pressure causes a slower horizon expansion, maintaining a relatively higher temperature at larger \( r_h \) values. These variations reveal how different physical factors influence temperature dynamics in BS growth.}
	\end{figure*}	
\section{Charged Black String}
In this section, we examine accretion with back-reaction effects for the charged BS. Cai and Zhang \cite{r15} derived cylindrically symmetric charged solutions to the EMFEs in the presence of a negative cosmological constant. The general form of the static charged BS metric in anti-de-Sitter space, where \(\alpha^{2} = -\frac{\Lambda}{3} > 0\), is given as,
\begin{align}
ds^{2}&=-\bigg(   \alpha^{2}r^{2}-\frac{4m_{0}}{\alpha r}+\frac{4q^{2}}{\alpha^{2}r^{2}}\bigg)  dt^{2}\nonumber\\
&+\frac{1}{\bigg(  \alpha^{2}r^{2}-\frac{4m_{0}}
	{\alpha r}+\frac{4q^{2}}{\alpha^{2}r^{2}}\bigg)}dr^{2}+r^{2}d\theta^{2}+\alpha^{2}r^{2}dz^{2}.
\label{044}
\end{align}
Note that the metric (\ref{044}) represents the solution at the zeroth-order approximation. To calculate the back-reaction, we use Eq. (\ref{1}). We find the perturbed EMFEs for Eq. (\ref{044}) using the same procedure as in the case of the static uncharged BS. The corrected line element for the charged BS is then given by,
\begin{align}
ds^{2}=&-\bigg(  \alpha^{2}r^{2}-\frac{4m(t,r)}{\alpha r}+\frac{4q^{2}}{\alpha^{2}r^{2}}\bigg)  dt^{2}\nonumber\\
&+\frac{1}{\bigg(\alpha^{2}r^{2}-\frac{4m(t,r)}
	{\alpha r}+\dfrac{4q^{2}}{\alpha^{2}r^{2}}\bigg)  }dr^{2}+r^{2}d\theta^{2}
+\alpha^{2}r^{2}dz^{2}.\label{045}
\end{align}
The components of the energy-momentum tensor for the charged BS are obtained by replacing,
\begin{equation}
\begin{array}
[c]{cc}%
T_{0}^{0}\longrightarrow T_{0}^{0}-\frac{q^{2}}{2\pi \alpha^{2}r^{4}}, &
T_{1}^{1}\longrightarrow T_{1}^{1}-\frac{q^{2}}{2\pi \alpha^{2}r^{4}},\\
T_{2}^{2}\longrightarrow T_{2}^{2}+\frac{q^{2}}{2\pi \alpha^{2}r^{4}}, &
T_{3}^{3}\longrightarrow T_{3}^{3}+\frac{q^{2}}{2\pi \alpha^{2}r^{4}}.
\end{array}
\label{046}
\end{equation}
Using the CAS Maple, we compute the components of the Einstein tensor, which are provided in Appendix A. The field equations for the charged BS are given in Appendix B. By solving them through successive integration and substitution, we obtain the correct mass, or running mass, for the charged BS, which is given by
\begin{equation}
4m(t,r)=-\int_{r_{0}}^{r}8\pi r^{2}T_{0}^{0}dr+\alpha^{2}r^{3}+m_{0}+8\pi
tr^{2}T_{0}^{1}.\label{047}%
\end{equation}
It is notable that Eq. (\ref{047}) aligns with Eq. (\ref{012}) in the first-order approximation, where charge-related terms involving \(q\) cancel out on both sides. This outcome is significant, as it indicates that, under the perturbative framework, the corrected mass \(m(t,r)\) remains consistent across both scenarios, unaffected by charge contributions.

Assuming that the energy-momentum tensor varies slowly with the radial coordinate, Eq. (\ref{047}) in the vicinity of the BS horizon,
\begin{equation}
4m(t,r)=-8\pi r^{2}(r-r_{0})|_{r=r_{0}}T_{0}^{0}+\alpha^{2}r^{3}+m_{0}+8\pi tr^{2}T_{0}^{1}.\label{048}%
\end{equation}
The above equation closely resembles Eq. (\ref{013}), further supporting the conclusion that charge does not influence the running mass in this perturbative scheme. The fact that the charge terms cancel out and do not affect the  mass \(m(t, r)\) underscores the robustness of the perturbative approach. It implies that the mass evolution near the BS horizon primarily depends on the matter content, as represented by the energy-momentum tensor, rather than on the presence of charge. This result shows that, for slowly varying energy-momentum tensors, the impact of charge on mass dynamics is minimal, suggesting that gravitational effects dominate the system's evolution over electromagnetic influences in this approximation.
\subsection{Corrected Apparent Horizon}
In order to discuss the thermodynamics in that case, first we need to find the corrected apparent horizon  for the charged BS.  From Eq. (\ref{045}) we have
\begin{equation}
\alpha^2 r^4 - 4m(t, r) r + 4q^2 = 0.
\end{equation}
The approximate solution is:
\begin{equation}
r_h \approx \left( \frac{4m(t, r)}{\alpha^2} \right)^{1/3} + \frac{q^2}{\alpha^2 \left( \frac{4m(t, r)}{\alpha^2} \right)^{5/3}}.
\end{equation}
This expression shows the leading term and a correction term due to the \( q^2 \)-charge term. The first term, $( \frac{4m(t, r)}{\alpha^2} )^{1/3} $, represents the dominant radius, while the second term is a small correction.
 
The charge \(q\) primarily contributes a stabilizing, outward-pushing correction to the apparent horizon, which is most significant at smaller radii (\(r\)). This effect counteracts gravitational collapse, slows down the evolution of the apparent horizon, and delays its convergence to the event horizon. In all three cases, the charge acts to reduce the effective gravitational pull, leading to an apparent horizon that is slightly larger and less dynamic than in uncharged cases. However, at larger radii or as \(t \to \infty\), the charge’s influence diminishes, and the accretion dynamics (mass, pressure, or flux) dominate the evolution of the horizons.

We plot the apparent horizon \( r_h \) of a charged BS over time \( t \) for varying charge \( q \) values. The charge affects the gravitational pull, influencing the horizon's evolution. In Fig. 5(a), \( r_h \) grows linearly with time, with a slight increase in growth rate for higher \( q \). The charge enhances gravitational attraction, accelerating accretion and horizon expansion. Unlike typical charged BS where electrostatic repulsion counteracts gravity, the dust model here produces a steady, near-uniform horizon growth.

Fig. 5(b) shows the perfect fluid model, where \( r_h \) increases non-linearly over time, especially for higher \( q \). The fluid’s pressure adds dynamics to the accretion rate, with charge significantly accelerating horizon growth as both electromagnetic and fluid pressures drive expansion.

In Fig. 5(c), the radiative fluid model depicts \( r_h \) rising sharply beyond a certain threshold, with high \( q \) yielding near-exponential horizon growth. Radiative fluids, with high-energy particles, show strong charge sensitivity: larger charges promote rapid expansion after an initial steady phase due to increased radiation pressure alongside gravitational attraction.
	\begin{figure*}[t]
		\centering
		\begin{minipage}[t]{0.38\textwidth}
			\centering
			\includegraphics[width=1\linewidth]{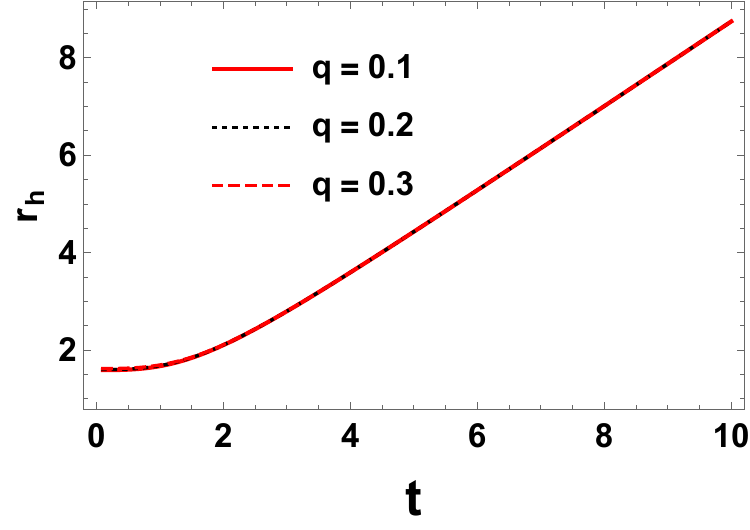}
			\caption*{Fig. 6(a)}
		\end{minipage}
		\hspace{1.8cm}
		\begin{minipage}[t]{0.38\textwidth}
			\centering
			\includegraphics[width=1\linewidth]{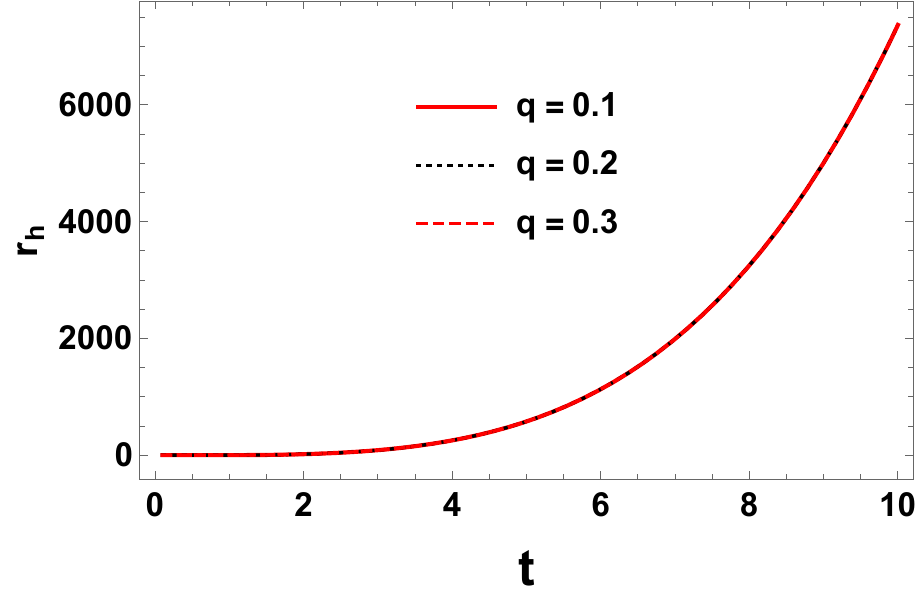}
			\caption*{Fig. 6(b)}
		\end{minipage}%
		~
		
		\vspace{.7cm}
		\begin{minipage}[t]{0.38\textwidth}
			\centering
			\includegraphics[width=1\linewidth]{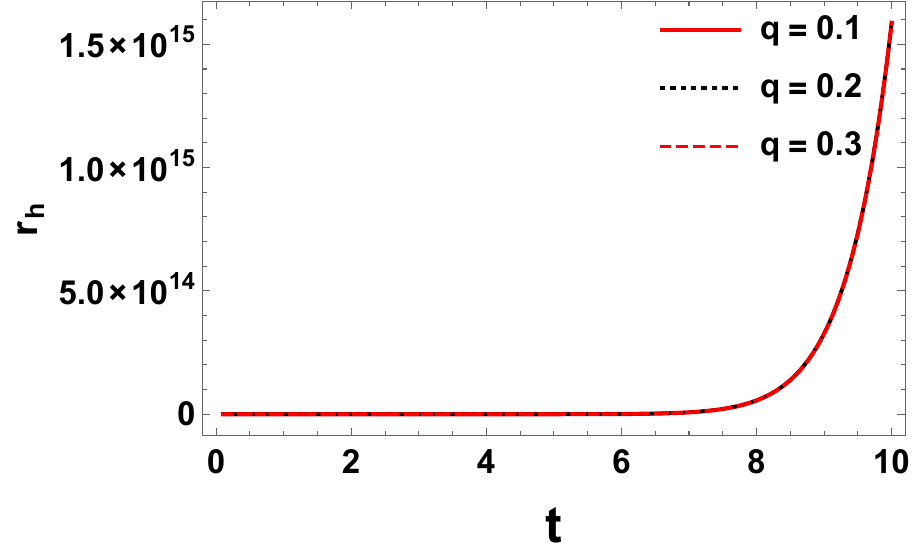}
			\caption*{Fig. 6(c)}
		\end{minipage}%
		\captionsetup{justification=raggedright,singlelinecheck=false}
	\caption{The corrected horizon \( r_h \) plotted against time \( t \) for various charge values \( q \) across different accretion models, highlighting the influence of charge on horizon growth. Fig. 6(a) illustrates the dust model, where the horizon grows linearly, and increasing charge causes moderate acceleration in growth. Fig. 6(b) displays the perfect fluid model, where the presence of both charge and pressure leads to non-linear horizon growth with significant acceleration. Fig. 6(c) represents the radiative fluid model, showing explosive horizon growth as \( q \) increases, driven by radiation pressure and charge effects. These results demonstrate that charge \( q \) has a progressively stronger impact on horizon growth across the models, with radiative fluids being most sensitive, reflecting the combined effects of charge and fluid type on accretion dynamics and horizon size in a charged BS scenario.}
	\end{figure*}	
	\subsection{Thermodynamics}
To discuss the thermodynamics, first need to identify the form the relevant parameters. The line element (\ref{045}) resembles that of a charged BH, likely a generalization of the Reissner-Nordström solution in a cylindrical symmetry. Using Eq. (\ref{040}), the surface gravity \(\kappa\) at the event horizon \(r = r_h\) is given by
\begin{equation}
\kappa = \frac{1}{2} \left| 2 \alpha^{2} r_h + \frac{4 m(t, r_h)}{r_h^{2}} - \frac{4}{r_h} \frac{\partial m(t, r)}{\partial r} \bigg|_{r = r_h} -\frac{8 q^{2}}{\alpha^{2} r_h^{3}} \right|.
\end{equation}
Consequently the the expression for corrected temperature is given as	
\begin{equation}
T = \frac{1}{4\pi} \left( 2\alpha^{2}r_h + \frac{4m_{0}}{r_H^2} - \frac{8q^{2}}{\alpha^{2}r_h^{3}} \right).
\end{equation}
This shows how mass and charge influence the thermal behavior of charge BS. The mass term reduces temperature as the corrected apparent horizon grows, while the charge introduces further cooling, stabilizing the BS. The scaling parameter \( \alpha \), modifying the temperature linearly with the horizon radius, showing back-reaction effects on its thermal properties.

To find the expression for entropy, use Eq. \ref{036} and \ref{037} we have,
\begin{equation}
S = \frac{A}{4} = \frac{2 \pi \alpha L \, r_h^2}{4} = \frac{\pi \alpha L \, r_h^2}{2}.
\end{equation}
This  expression is similar to Eq. \ref{0040} and shows the entropy depends on the horizon radius \(r_h\).

\section{Conclusion and Discussion}

The study examines the back-reaction effects of accreting fluid on a BS using a perturbative approach. Two primary methods are typically considered in accretion analysis. The first neglects back-reaction, which is valid when the mass of the accreting matter is negligible compared to that of the BH. The second approach includes the back-reaction, leading to complete solutions to the EFEs and the matter equations. In this work, we consider a  BS, applying a perturbative scheme to approximate the EFEs and derive an expression for the corrected mass of the BS near the horizon, as shown in Eq. (\ref{013}).

We further analyzed energy conditions for our solution, demonstrating that it satisfies all energy conditions provided that the running mass \(m\) fulfills the condition \( m' \leq 3\alpha^{2} r^{2} \), where the prime denotes differentiation with respect to \(r\). This indicates that both energy conditions and EFEs are satisfied, confirming the self-consistency of our solution.

Accretion with back-reaction was explored for three matter models: dust, perfect fluid, and radiative fluid. In each case, we showed graphically that the running or corrected mass \(m(t, r)\) increases over time \(t\) and with radial distance \(r\), emphasizing the impact  of back-reaction in each model. The dust model shows steady, linear mass growth, especially in outer regions. The perfect fluid model exhibits nonlinear accretion due to pressure effects, while the radiative fluid model accelerates mass accumulation at larger radii due to radiation pressure. In central regions, differences among models are less pronounced, as gravitational effects dominate uniformly. Overall, outer regions display faster mass increase, driven by fluid-specific dynamics and accretion properties
This is depicted in Fig. 1(a)-(f).

We used the corrected metrics given by Eq. (\ref{3}) and (\ref{013}) to calculate the corrected apparent horizon due to the back-reaction effect. We have demonstrated that the corrected apparent horizon \( r_h \) varies with time \( t \) and energy density \( \rho \) or pressure \( P \) for the dust, perfect fluid, and radiative fluid models in accretion. In the dust model, which has no internal pressure, accretion occurs steadily. In contrast, the perfect fluid model shows a steeper horizon growth due to pressure resistance. For the radiative fluid case, radiation pressure can either accelerate or stabilize horizon expansion, depending on the balance with gravitational forces. The These are illustrated graphically in Fig. 3(a)-(d), respectively. We present the Carter-Penrose diagram to illustrate the causal structure and the relationship between the corrected apparent horizon and the event horizon for all three distinct cases. Our analysis shows that the accretion of matter plays a crucial role in modifying the spacetime geometry. Specifically, corrections arising from dust, radiative fluid, and the energy density and pressure of a perfect fluid alter the location of the apparent horizon. Furthermore, we discuss the thermodynamics of the corrected line element. Using the corrected apparent horizon, we calculate and plot the expressions for entropy and temperature for each model. The increase in entropy across all models is consistent with thermodynamic principles, while rapid horizon growth leads to a sharp drop in temperature. The graphs of entropy and temperature for all three cases are shown in Figs. 4 and 5, respectively.

We also extended our analysis to the charged BS and applied the perturbation technique to examine its properties. We found that the expression for the running mass \(m(t, r)\) of the charged BS is identical to that of the uncharged BS, suggesting that charge does not influence the mass in this perturbative technique. Furthermore, we derived the corrected expression for the apparent horizon in the charged case and explored the impact of different charge values \(q\) on this correction. Additionally, we investigated the thermodynamic properties of the charged BS. While the entropy remained the same for both charged and uncharged cases, we observed a slight variation in the temperature due to the different surface gravities in each case.

The running mass \( m(t, r) \) is associated with the asymptotically flat ADM (Arnowitt-Deser-Misner) mass, which measures the total mass-energy content of a spacetime. The ADM mass incorporates both gravitational energy and matter fields. For a spacetime to be asymptotically flat, the metric must approach the Minkowski metric at spatial infinity. In general, the ADM mass can be expressed in terms of the stress-energy tensor \( T^{\mu}_{\nu} \). For a spherically symmetric distribution of matter, the ADM mass is given by
\begin{equation}
M_{\text{ADM}} = \lim_{r \to \infty} m(t, r).
\end{equation}
The asymptotic behavior of the mass function relates to the ADM mass as \( r \to \infty \). If the energy density \( T_0^0 \) and pressure \( T_0^1 \) vanish or become negligible at infinity, the mass function approaches a constant value. At large \( r \), the dominant contributions typically arise from the \( m_0 \) term and the \( \alpha^2 r^3 \) term, assuming the integral converges.
For \( m(t, r) \) to represent the ADM mass, the spacetime must satisfy the condition of asymptotic flatness, as given,
$$
M_{\text{ADM}} = m_0 \quad (\text{assuming other terms vanish as } r \to \infty).
$$

It is also worthwhile to consider that the perturbative technique can be extended to rotating BS and rotating charged  BS. The perturbed solutions, along with the effects of various energy-momentum tensors on the corrected masses, will be of significant interest.

\appendix
\section*{Appendix A: Einstein Tensor Components}
The Einstein tensors for the considered spacetime are given by:
\begin{align}
G_{0}^{0} &= G_{1}^{1} = \frac{-1}{\alpha^{2} r^{4}}( -3\alpha^{4} r^{4} + 4\alpha^{2} r^{2}  m' + 4q^{2})\\
 G_{0}^{1}& = \frac{4\dot{m}}{r^{2}}   ,~G_{1}^{0} = \frac{-4 \alpha^{4} r^{2} \dot{m} }{(-\alpha^{4} r^{4} + 4m \alpha^{2} r - 4q^{2})^{2}}
\end{align}
\begin{align}
G_{2}^{2} &= G_{3}^{3} =  \frac{1}{\alpha^{2} r^{4} ( -\alpha^{4} r^{4} + 4m \alpha^{2} r - 4q^{2} )^{3}} [ 2\alpha^{10} r^{11} \ddot{m}\nonumber\\
& + 2\alpha^{14} r^{15} m'' - 384\alpha^{4} r^{4} q^{6} 
- 384\alpha^{4} r^{4} m q^{4} m''\nonumber\\
& + 96\alpha^{6} r^{7} q^{4} m'' - 128\alpha^{8} r^{6} m^{3} m'' - 8\alpha^{8} r^{8} \ddot{m} m \nonumber\\
&+ 8\alpha^{6} r^{7} \ddot{m} q^{2} - 24\alpha^{12} r^{12} m m'' + 24\alpha^{10} r^{11} q^{2} m''\nonumber\\  
&  + 96\alpha^{10} r^{9} m^{2} m'' - 192\alpha^{8} r^{8} m q^{2} m'' + 384\alpha^{6} r^{5} m^{2} q^{2} m''\nonumber\\ 
&- 256 q^{8} - 3\alpha^{16} r^{16} + 128\alpha^{2} r^{3} q^{6} m'' - 192\alpha^{8} r^{8} q^{4}\nonumber\\
& + 192\alpha^{10} r^{7} m^{3} - 144\alpha^{12} r^{10} m^{2}- 40\alpha^{12} r^{12} q^{2} \nonumber\\
&+ 36\alpha^{14} r^{13} m + 16\alpha^{8} r^{8} t( \dot{m} )^{2} + 960\alpha^{6} r^{5} m q^{4}\nonumber\\
& + 256\alpha^{6} r^{3} m^{3} q^{2} - 768\alpha^{4} r^{2} m^{2} q^{4} + 336\alpha^{10} r^{9} m q^{2}\nonumber\\
& + 768\alpha^{2} r m q^{6} - 768\alpha^{8} r^{6} m^{2} q^{2}].
\end{align}
\section*{Appendix B: Energy-Momentum Tensor Components}
In this appendix, we present the components of the energy-momentum tensor \( T^{\mu}_{\nu} \). The following relations are derived from the EFEs. The energy-momentum tensor components are given by:
\begin{align}
&8\pi \bigg( T_{0}^{0}- \frac{q^{2}}{2\pi \alpha^{2} r^{4}} \bigg) =8\pi \bigg( T_{1}^{1} - \frac{q^{2}}{2\pi \alpha^{2} r^{4}} \bigg)\\
 &=-\frac{-3\alpha^{4} r^{4} + 4\alpha^{2} r^{2} m' - 4q^{2}}{\alpha^{2} r^{4}}, \label{d7} \nonumber\\
&8\pi T_{0}^{1} = \frac{4 \dot{m}}{r^{2}}, 8\pi T_{1}^{0} = \frac{-4\alpha^{4} r^{2} \dot{m}}{( -\alpha^{4} r^{4} + 4\alpha^{2} r m - 4q^{2})^{2}}, 
\end{align}
\begin{align}
&8\pi \left[ T_{3}^{3} + \frac{q^{2}}{2\pi \alpha^{2} r^{4}} \right]=8\pi \left[ T_{2}^{2} + \frac{q^{2}}{2\pi \alpha^{2} r^{4}} \right]\nonumber\\
&=\frac{1}{\alpha^{2} r^{4} ( -\alpha^{4} r^{4} + 4m \alpha^{2} r - 4q^{2} )^{3}} [ 2\alpha^{10} r^{11} \ddot{m}\nonumber\\
& + 2\alpha^{14} r^{15} m'' - 384\alpha^{4} r^{4} q^{6} 
- 384\alpha^{4} r^{4} m q^{4} m''\nonumber\\
& + 96\alpha^{6} r^{7} q^{4} m'' - 128\alpha^{8} r^{6} m^{3} m'' - 8\alpha^{8} r^{8} \ddot{m} m \nonumber\\
&+ 8\alpha^{6} r^{7} \ddot{m} q^{2} - 24\alpha^{12} r^{12} m m'' + 24\alpha^{10} r^{11} q^{2} m''\nonumber\\  
&  + 96\alpha^{10} r^{9} m^{2} m'' - 192\alpha^{8} r^{8} m q^{2} m'' + 384\alpha^{6} r^{5} m^{2} q^{2} m''\nonumber\\ 
&- 256 q^{8} - 3\alpha^{16} r^{16} + 128\alpha^{2} r^{3} q^{6} m'' - 192\alpha^{8} r^{8} q^{4}\nonumber\\
& + 192\alpha^{10} r^{7} m^{3} - 144\alpha^{12} r^{10} m^{2}- 40\alpha^{12} r^{12} q^{2} \nonumber\\
&+ 36\alpha^{14} r^{13} m + 16\alpha^{8} r^{8} t( \dot{m} )^{2} + 960\alpha^{6} r^{5} m q^{4}\nonumber\\
& + 256\alpha^{6} r^{3} m^{3} q^{2} - 768\alpha^{4} r^{2} m^{2} q^{4} + 336\alpha^{10} r^{9} m q^{2}\nonumber\\
& + 768\alpha^{2} r m q^{6} - 768\alpha^{8} r^{6} m^{2} q^{2}].
\end{align} \label{d11b}


\begin{thebibliography}{99}                         
	\bibitem{rl} Lemos, J. P. S. “Cylindrical Black Hole in General Relativity.” \textit{Physics Letters B} 353, no. 1 (1995): 46.
	
	\bibitem{2} Cai, R. G., and Y. Zhang. “Black Plane Solutions in Four Dimensional Spacetimes.” \textit{Physical Review D} 54, no. 8 (1996): 4891.
	
	\bibitem{3} Lemos, J. P. S., and V. T. Zanchin. “Rotating Charged Black Strings and Three-Dimensional Black Holes.” \textit{Physical Review D} 54, no. 6 (1996): 3840–3853.
	
	\bibitem{4} Bondi, H. “On Spherically Symmetrical Accretion.” \textit{Monthly Notices of the Royal Astronomical Society} 112, no. 1 (1952): 195.
	
	\bibitem{r5} Lynden-Bell, D. “Galactic Nuclei as Collapsed Old Quasars.” \textit{Nature} 223, no. 5207 (1969): 690.
	
	\bibitem{r6} Shakura, N. I., and R. A. Sunyaev. “Black Holes in Binary Systems: Observational Appearance.” \textit{Astronomy and Astrophysics} 24, no. 1 (1973): 337–355.
	
	\bibitem{r7} Pringle, J. E., and M. J. Rees. “Accretion Disc Models for Compact X-Ray Sources.” \textit{Astronomy and Astrophysics} 21, no. 1 (1972): 1–9.
	
	\bibitem{r8} Salpeter, E. E. “Accretion of Interstellar Matter by Massive Objects.” \textit{Astrophysical Journal} 140, no. 2 (1964): 796–800.
	
	\bibitem{r9} Thorne, K. S. “Disk-Accretion onto a Black Hole. II. The Tidal, Viscous, and Thermal Processes.” \textit{Astrophysical Journal} 191, no. 1 (1974): 507–520.
	
	\bibitem{r10} Bardeen, J. M., and R. V. Wagoner. “Relativistic Disks in the Electromagnetic Field of a Black Hole.” \textit{Astrophysical Journal} 167, no. 1 (1971): 7–18.
	
	\bibitem{r11} Bardeen, J. M., and R. V. Wagoner. “The Effect of the Accretion Process on the Properties of Black Holes.” \textit{Astrophysical Journal} 243, no. 1 (1981): 257–262.
	
	\bibitem{r12} Krolik, J. H. \textit{Active Galactic Nuclei: From the Central Black Hole to the Galactic Environment}. Princeton: Princeton University Press, 1999.
	
	\bibitem{r13} Barack, L., and A. Pound. “Self-Force and Radiation Reaction in General Relativity.” \textit{Reports on Progress in Physics} 82, no. 1 (2019): 016904.
	
	\bibitem{r14} Poisson, E. \textit{A Relativist's Toolkit: The Mathematics of Black-Hole Mechanics}. Cambridge: Cambridge University Press, 2004.
	
	\bibitem{r15} Babichev, E., M. Charbit, and S. Ramazanov. “Accretion onto Black Holes in Extended Gravity Models.” \textit{Classical and Quantum Gravity} 35, no. 15 (2018): 154002. DOI: 10.1088/1361-6382/aac9f4.
	
	\bibitem{r16} Babichev, E., V. Dokuchaev, and Y. Eroshenko. “Black Hole Mass Decreasing Due to Phantom Energy Accretion.” \textit{Physical Review Letters} 93, no. 2 (2004): 021102. DOI: 10.1103/PhysRevLett.93.021102.
	
	\bibitem{r17} Blandford, R. D., and R. L. Znajek. “Electromagnetic Extraction of Energy from Kerr Black Holes.” \textit{Monthly Notices of the Royal Astronomical Society} 179, no. 3 (1977): 433–456.
	
	\bibitem{r18} Karkowski, J., and E. Malec. “Backreaction of Accreting Matter on the Schwarzschild Black Hole.” \textit{Physical Review D} 70, no. 4 (2004): 044006.
	
	\bibitem{r19} Barausse, E., and L. Rezzolla. “Accretion onto a Black Hole in an Asymptotically Flat Spacetime: Interaction of the Inner and Outer Hydrodynamic Flows.” \textit{Physical Review D} 77, no. 10 (2008): 104027.
	
	\bibitem{r20} Vaidya, P. C. “Newtonian Time in General Relativity.” \textit{Nature} 171, no. 4340 (1953): 260.
	
	\bibitem{r21} Tolman, R. C. “Effect of Inhomogeneity on Cosmological Models.” \textit{Proceedings of the National Academy of Sciences} 20, no. 3 (1934): 169.
	
	\bibitem{r22} Babichev, E., V. Dokuchaev, and Y. Eroshenko. “Back-Reaction of Accreting Matter Onto a Black Hole in the Eddington–Finkelstein Coordinates.” \textit{Classical and Quantum Gravity} 29, no. 11 (2012): 115002.
	
	\bibitem{r23} Bañados, M., C. Teitelboim, and J. Zanelli. “The Black Hole in Three-Dimensional Space-Time.” \textit{Physical Review Letters} 69, no. 13 (1992): 1849–1851. DOI: 10.1103/PhysRevLett.69.1849.
	
	\bibitem{r24} Misner, C. W., K. S. Thorne, and J. A. Wheeler. \textit{Gravitation}. San Francisco: W. H. Freeman, 1973.
	
	\bibitem{r25} Geroch, R., and P. S. Jang. “Motion of a Body in General Relativity.” \textit{Journal of Mathematical Physics} 16, no. 1 (1975): 65–67.
	
	\bibitem{r26} Price, R. H. “Nonspherical Perturbations of Relativistic Gravitational Collapse. I. Scalar and Gravitational Perturbations.” \textit{Physical Review D} 5, no. 10 (1972): 2419–2438.
	
	\bibitem{r27} Bondi, H. “On Spherically Symmetrical Accretion.” \textit{Monthly Notices of the Royal Astronomical Society} 112, no. 1 (1952): 195.
	
	\bibitem{r28} Michel, F. C. “Accretion of Matter by Condensed Objects.” \textit{Astrophysics and Space Science} 15, no. 1 (1972): 153.
	
	\bibitem{r29} Babichev, E., V. Dokuchaev, and Y. Eroshenko. “Black Hole Mass Decreasing Due to Phantom Energy Accretion.” \textit{Physical Review Letters} 93, no. 2 (2004): 021102. DOI: 10.1103/PhysRevLett.93.021102.
	
	\bibitem{r30} Shapiro, S. L., and S. A. Teukolsky. \textit{Black Holes, White Dwarfs, and Neutron Stars: The Physics of Compact Objects}. New York: Wiley-Interscience, 1983.
	
	\bibitem{r31} Hawking, S. W., and G. F. R. Ellis. \textit{The Large Scale Structure of Space-Time}. Cambridge: Cambridge University Press, 1973.
	
	\bibitem{r32} Rizwan, M. ``Corrected Entropy Law for Charged and Rotating Black Strings.'' \textit{International Journal of Theoretical Physics} 55 (2016): 3515--3523.
	
	
	\bibitem{r33} Bekenstein, J. D. “Black Holes and Entropy.” \textit{Physical Review D} 7, no. 8 (1973): 2333–2346.
	
\end{thebibliography}
\end{document}